\documentclass{article}

\usepackage{arxiv}

\usepackage[utf8]{inputenc} 
\usepackage[T1]{fontenc}    
\usepackage{hyperref}       
\usepackage{url}            
\usepackage{booktabs}       
\usepackage{amsfonts}       
\usepackage{nicefrac}       
\usepackage{microtype}      
\usepackage{lipsum}

\usepackage{amsmath,dsfont,mathrsfs,color,verbatim}
\usepackage[numbers]{natbib}
\usepackage{multicol}
\usepackage{multirow}
\usepackage{float}
\usepackage{soul}
\usepackage{cancel}
\usepackage{graphicx,amssymb,subfig,icomma}

\newcommand\sbullet[1][.5]{\mathbin{\vcenter{\hbox{\scalebox{#1}{$\bullet$}}}}}

\title{Large-data determinantal clustering}

\author{
  Serge Vicente \\
  Department of Mathematics and Statistics\\
  Université de Montréal\\
  Montréal, Québec, Canada \\
  \texttt{s.vicente@umontreal.ca} \\
   \And
 Alejandro Murua \\
  Department of Mathematics and Statistics\\
  Université de Montréal\\
  Montréal, Québec, Canada \\
  \texttt{alejandro.murua@umontreal.ca} \\
}

\begin{document}
\maketitle

\begin{abstract}
Determinantal consensus clustering is a
  promising and attractive alternative to partitioning about medoids and $k$-means
  for ensemble clustering.
  Based on a determinantal point process or DPP sampling, it ensures that subsets of similar points are less likely to be selected as centroids. It favors
  more diverse subsets of points.
  The sampling algorithm of the determinantal point process requires the eigendecomposition of a
  Gram matrix. This becomes computationally intensive when the data size is very large.
  This is particularly an issue in consensus clustering, where a given clustering algorithm is run
  several times in order to produce a final consolidated clustering.
  We propose two efficient alternatives to carry out determinantal consensus clustering on large datasets.
  They consist in DPP sampling based on sparse and small kernel matrices whose eigenvalue distributions
  are close to that of the original Gram matrix.
\end{abstract}


\section{Introduction}
Cluster analysis is a classical procedure to aggregate elements according to their similarity.
Among the most popular methods to do clustering, we find
the well-known $k$-means
  \citep{Lloyd19823} and the partitioning around medoids (PAM) \citep{Kaufmann19873} algorithms.
These two procedures have the particularity of
building data partitions from an initial choice of random points called centroids.
These points are usually sampled uniformly at random from the set of all datapoints.
Since each point has the same probability of being chosen,
the initial sample may end up with a set of points containing many similar points
that carry the same type of information. That is, the initial sample might not represent the diversity present in the data.
This might affect the effectiveness of the clustering.

Nowadays, the diversity in the selected elements is a major concern in some domains of research,
like clinical trials \citep{Clark2019}, forensic sciences \citep{Wagstaff2018} or educational development \citep{Szelei2019}.
Adopting uniform random sampling as a sampling mechanism can result in sets of elements with a poor coverage of all the facets of a population under study.
Determinantal point processes, or DPPs for short, introduced by \cite{Borodin200031},
can address this problem. DPPs model negative correlations between points so that
similar elements have less chances of being simultaneously sampled.
The negative correlations are captured by the so-called kernel or Gram matrix \citep{Kulesza20123}, a
matrix whose entries represent a measure of similarity between pair of points.
DPPs have already been adopted in machine learning as models for subset selection
  \citep{Hafiz20133,Gartrell2018,Shah2013,Gillenwater2012,Mariet2019}.

The origins of DPPs can be found in quantum physics \citep{Macchi19753}.
Known first as \textit{fermion processes,} they model the distribution of fermion systems at thermal equilibrium.
Much later, \cite{Borodin200031} introduced  the now accepted \textit{Determinantal Point Process} terminolgy in the
mathematics community.
DPPs have also been applied to problems dealing with nonintersecting random paths \citep{Daley20033},
random spanning trees \citep{Borodin20032}, and the study of eigenvalues of random matrices \citep{BenHough20063}.

Clustering algorithms like $k$-means and the partitioning around medoids result in single partition of data, seeking to maximize intra-cluster similarity and inter-cluster dissimilarity.
However, the clustering results of two different algorithms can be very different.
The lack of an external objective and impartial criterion can explain those differences \citep{Vega20113}.
The dependence on the initial choice of centroids is also an important factor.
\cite{Blatt19963} and \cite{Blatt19973} proposed a new approach to improve the quality and robustness of clustering results.
The approach was later formalized by \cite{Strehl20023}, where the notion of \emph{cluster ensembles} is introduced.
Cluster ensembles combine different data partitions into a single consolidated clustering.
A particular cluster ensemble method was later introduced in \cite{Monti20033}, the so-called \emph{consensus clustering}.
This method consists in performing multiple runs of the same clustering algorithm on the same data,
to produce a single clustering configuration by agreement among all the runs.

Most often, the particular clustering algorithm to be run several times for consensus implies random initial conditions or centroids.
Recently \cite{vicente&murua1-2020} introduced one such method, the \textit{determinantal consensus clustering} or
\textit{consensus DPP} procedure.
This generates centroids through a DPP process.
The similarity or distance between the datapoints is incorporated in the similarity matrix, also known as the kernel
or Gram matrix,
which constitutes the core of the DPP process. Moreover, the link between similarity matrices and kernel methods for statistical or machine learning makes the method very flexible and effective at discovering data clusters.
The diversity within centroids is automatically inherited in the DPP sampling.
The DPP ``diversity at sampling'' property has shown to greatly improve
the consensus clustering results \citep{vicente&murua1-2020}.

In practice, the centroids are drawn using the DPP sampling algorithm described in \cite{BenHough20063} and \cite{Kulesza20123}.
This algorithm is based on computing the spectral decomposition of the data similarity or kernel matrix.
Unfortunately, when the data size $n$ is very large,
the eigendecomposition becomes a computational burden.
The computational complexity of the eigendecomposition of a $n\times n$ symmetric matrix is of order
$O\left(n^3\right)$.
However, to sample the centroid points,
we might not need to compute all eigenvalues and eigenvectors of the kernel matrix.
In fact, the probability of selecting each datapoint depends on a
corresponding eigenvalue.
Datapoints associated with relatively very small eigenvalues are selected with very low probability.
Therefore, the key to reduce the computational burden induced by large datasets is
the extraction of the largest eigenvalues of the associated kernel matrix.
This raises the necessity to deliver algorithms able to extract such a subset of eigenvalues.
One of the most used algorithms to extract the largest eigenvalues is the Lanczos algorithm \citep{Lanczos1950}.
Due to his proven numerical instability, many variations of it have been proposed.
A popular variation of the Lanczos algorithm is the \textit{implicitly restarted Lanczos method} \citep{Calvetti1994}
which we adopte in this paper.

The Lanczos algorithm have been specially developed for large sparse and symmetric matrices.
Hence, to be able to perform determinantal consensus clustering on large datasets,
we need to find good sparse approximations of the often dense \textit{original} kernel matrix.
We propose two sound approaches: one approach based on the Nearest Neighbor
Gaussian Process of \cite{Datta2016}, and another approach based on a
random sampling of small submatrices from the dense kernel matrix. This latter approach may be seen
as a kind of divide and conquer approach.
Although we show that these approaches offer good approximations to the eigenvalue distribution of the
original kernel matrix, our goal is  rather to introduce alternative efficient DPP sampling models
for large datasets
that inherit the data diversity
expressed in the original kernel matrix.

The paper is organized as follows: in Section~\ref{sec:determinantal:consensus},
we summarize the consensus clustering  and recall
the basic characteristics of determinantal point processes.
The determinantal consensus clustering is also summarized in this section.
In Section~\ref{largedatasets}, we introduce the problem of using the determinantal point process
in the context of large datasets, and present two approaches to address this issue.
In Section~\ref{experiments2}, we evaluate the two approaches presented in Section~\ref{largedatasets}
through large dataset simulations; here we also
illustrate the concept of diversity introduced by the determinantal point process.
A  performance comparison between our two approaches and two other competing methods
on large real datasets is shown in Section~\ref{sec:real}.
We conclude with a few thoughts and a discussion in Section~\ref{sec:conclusions}.

\section{Determinantal consensus clustering}\label{sec:determinantal:consensus}

\subsection{Consensus clustering}\label{sec:consensus}
Throughout the paper, the data will be
denoted by $\mathcal{S}=\{{x_1}, \dots, {x_n}\}\subset \mathds{R}^p$, where $x_i$ represents a $p$-dimensional vector, for $i=1,\dots,n$ and $n \geq 2$. Consider
 a particular clustering algorithm run $R$ times on the same data $\mathcal{S}$. The agreement among the several runs of the algorithm is based on the \emph{consensus matrix} $C$. This is a $n\times n$ symmetric matrix whose entries $\{ C_{ij}, \; i, j=1,\dots, n\}$ represent the proportion of runs in which elements $x_i$ and $x_j$ of $\mathcal{S}$ fall in the same cluster.
 Let $r$ represent a specific run of the clustering algorithm, $r=1,\dots, R,$ and let $C_r$ be the associated $n\times n$ symmetric binary matrix with entries $c^{r}_{ij}= 1$ if  $x_i$ and $x_j$ belong to the same cluster in the $r$th run of the algorithm, and
 $c^{r}_{ij}= 0$, otherwise,
 $i, j=1,\dots, n$.
 The components of the consensus matrix $C$ are given by $C_{ij}=\sum_{r=1}^{R}c^{r}_{ij} / R,$
 $i, j=1,\dots,n$. The entry $C_{ij}$ is known as \emph{consensus index}.
 The diagonal entries are given by $C_{ii}=1,$ for $i=1,\dots, n$.

 Our interest is to extend the determinantal consensus clustering (or consensus DPP) algorithm of \cite{vicente&murua1-2020} to large datasets. Consensus DPP is a modified version of Partitioning around medoids (PAM) algorithm \citep{Kaufmann19873}
 in which the center points are sampled with a determinantal point process (DPP); more details on DPP are shown in the next section.
 Each run starts with a Voronoi diagram \citep{Aurenhammer19913} on the set $\mathcal{S}$. This partitions the space into several cells or regions based on a random subset of {\it generator points.}
 After $R$ runs of the algorithm, we obtain $R$ partitions associated with the $R$ Voronoi diagrams.
 The consensus matrix $C$ is computed from these partitions.
 Let $\theta \in [0,1]$ be a proportion threshold.
 According to \cite{Blatt19963}, if $C_{ij}\geq \theta$, points $x_i$ and $x_j$ are defined as ``friends'' and are included in the same consensus cluster.
 Moreover, all mutual friends (including friends of friends, etc.) are assigned to the same cluster.
 To select an appropriate threshold value, we follow \cite{Murua20143} and consider all threshold values
 given by the set of all different observed consensus indexes $C_{ij}$.
 If there are $t$ different consensus indexes, we will have a collection of $t$ thresholds $\theta_1, \theta_2, \dots, \theta_t$.
 For each threshold $\theta_i, \; i=1,\dots, t$, a consensus clustering configuration with $K(\theta_i)$ clusters is obtained.
 If $\theta_i=0$, we obtain a configuration with $K(0)=1$ cluster, that is, a single cluster equal to $\mathcal{S}$.
 If $\theta_i=1$, we obtain a configuration with $K(1) =n$ singleton clusters; that is, each element of $\mathcal{S}$ forms a singleton cluster.
 In general, clustering configurations with one cluster or $n$ clusters are of no interest. Therefore,
 thresholds $\theta_i$ that are too low or too large are not relevant.
 This observation leads to a more efficient procedure \citep{vicente&murua1-2020}
 where only a predetermined sequence of $t$ thresholds $\tau < \theta_1 < \theta_2 < \cdots <\theta_t$,
 bounded from below by a certain minimum threshold $\tau$, are considered.
 This approach is particularly useful in the context of large datasets, since reducing the range of considered thresholds can reduce the computational burden induced by the high number of elements.

 Moreover, we are not interested in a clustering configuration with too many small clusters.
 Only clustering configurations with cluster sizes larger than a minimal value are admissible (that is, accepted).
 \cite{vicente&murua1-2020} established through extensive simulations that
 the ``square-root choice'' (for the number of bins of a histogram) $\sqrt{n}$ is an adequate value for the minimum cluster size.
 Each one of the $t$ consensus clustering configurations obtained with the $t$ thresholds $\{\theta_i\}$ are examined
to verify that they are admissible.
For every non-admissible consensus clustering configuration
we merge each small cluster with its closest ``large'' cluster, according to the following procedure, inspired by single linkage:
(i) select the cluster $\mathcal{V}$ that has the smallest cluster size $< \sqrt{n}$;
(ii) find the pair of indexes $(i^*, j^*) \in \{1,\ldots, n\}$ that satisfies
  $C_{i^*  j^*} = \max\{ C_{ij} : x_i \in \mathcal{V}, x_j \not\in \mathcal{V} \}$;
(iii)  merge the cluster $\mathcal{V}$ to the cluster that includes  $x_{j^*}$;
(iv) repeat the merging procedure until there are no more clusters with size  smaller than $\sqrt{n}$.
Once all $t$ consensus clustering configurations are made admissible, we proceed to select the {\it final consensus clustering}
as the consensus clustering configuration among these $t$ configurations that minimizes the kernel-based validation index
of \cite{Fan20103}. This index was conceived after the studies of \cite{Girolami20023} to choose the optimal final cluster among several possible partitions. It can be seen as an index that combines modified extensions of the
between and within variances to kernel-based methods.
See \cite{Fan20103} or \cite{vicente&murua1-2020} for further details.

\subsection{The determinantal point process}

Let $L$ be a $n\times n$ real symmetric and positive semidefinite matrix that measures similarity between all pairs of elements of $\mathcal{S}$.
We denote by $L_Y=\left(L_{ij}\right)_{i,j\in Y}$
the principal submatrix of $L$ whose rows and columns are indexed by
the subset $Y\subseteq \mathcal{S}$.
A determinantal point process, DPP for short, is a probability measure on $2^{\mathcal{S}}$ that assigns probability
\begin{equation}\label{definition2}
  P\left( Y \right)= \det(L_Y) / \det(L+I_n),
\end{equation}
to any subset $Y\in 2^{\mathcal{S}}$,
where $I_n$ is the identity matrix of dimension $n \times n$.
We write $\boldsymbol{Y}\sim DPP_{\mathcal{S}}(L)$ for the corresponding determinantal process.

The matrix $L$ is known as the kernel matrix of the DPP \citep{Kulesza20123,Kang20133,Hafiz20143}.
It can be shown \citep{Kulesza20123} that $\det(L+I_n) = \sum_{Y \subset 2^\mathcal{S}} \det(L_Y)$, hence
\eqref{definition2} does indeed define a probability mass function over all subsets in $2^\mathcal{S}$.
This definition states restrictions on all the principal minors of the kernel matrix $L$, denoted by $\det(L_Y)$.
Indeed, as $P\left(\boldsymbol{Y}=Y\right)\propto\det(L_Y)$ represents a probability measure, we have $\det(L_Y)\geq0$, for any $Y\subseteq \mathcal{S}$.

Any symmetric positive semidefinite matrix $L$ may be a kernel matrix of a DPP.
For the construction of the similarity matrix $L$ in \eqref{definition2}, we use a suitable Mercer Kernel \citep{Girolami20023}
as indicated in \cite{vicente&murua1-2020}.
The choice of an appropriate kernel is a critical step in the application of any kernel-based method.
However, as pointed out by \cite{Howley20063}, there is no rule nor consensus about its choice.
Ideally, the kernel must be chosen according to prior knowledge of the problem domain
\citep{Howley20063,Lanckriet20043}, but this practice is rarely observed.
In the absence of expert knowledge, a common choice is the
\emph{Radial Basis Function} (RBF) kernel (or \emph{Gaussian kernel}):
\begin{equation*} 
  L=\bigl( \exp\bigl\{ -\|x_i-x_j\|^2 / (2\sigma^2) \bigr\}\bigr)_{i,j=1}^n,
\end{equation*}
where the scale parameter $\sigma$, known as the kernel's bandwidth, represents the relative spread of the Euclidean distances $\|x_i-x_j\|$ between any two points $x_i$ and $x_j$.
Due to its appealing mathematical properties, this particular kernel has been extensively used in many studies.
A particular property of the Gaussian kernel is that it is positive and bounded from above by one,
 making it directly interpretable as a scaled measure of similarity between any given pair of points.

The computation of the RBF kernel requires the estimation of the bandwidth parameter $\sigma$. As pointed by \cite{Murua20143}, most of the literature considers $\sigma$ as a parameter that can be estimated by observed data.  Inspired by \cite{Blatt19963} and \cite{Blatt19973}, we estimate $\sigma^2$ by the average of all pairwise and squared Euclidean distances, i.e.,
$  \widehat{\sigma}^2= 2 \sum_{i<j}\|(x_i-x_j)\|^2 / \bigl( n(n-1)\bigr).$ Other choices of estimating $\sigma^2$ are referred in \cite{vicente&murua1-2020}. We chose the average for its simplicity and fast computation even when the dataset is large.

\section{The case of large datasets}\label{largedatasets}

\cite{BenHough20063}
and \cite{Kulesza20123} present an efficient scheme to sample from a DPP.
The algorithm is based on the following observations.
Let $L=\sum_{i=1}^{n} \lambda_i(L){v}_i{v}_i^T$ be an orthonormal eigendecomposition of $L$,
where $\lambda_1(L) \geq \lambda_2(L) \geq \cdots \geq \lambda_n(L) \geq 0$ are the eigenvalues of $L$,
and $\{v_i : i=1,\ldots, n\}$ are the eigenvectors of $L$.
For any set of indexes $J \subseteq \{1,2,\ldots, n\}$, define the subset of eigenvectors $V_J = \{ v_i : i \in J\}$,
and the associated matrix $\mathcal{K}_J = \sum_{i\in J} {v}_i {v}_i^T.$
It can be shown that the matrix $\mathcal{K}_J$ defines a so-called \emph{elementary} DPP which we denote by DPP$(\mathcal{K}_J)$.
It turns out that the DPP$_\mathcal{S}(L)$ is a mixture of all elementary DPP given by the index sets $J$.
That is
\[
\operatorname{DPP}_\mathcal{S}(L) = \sum_{J} \operatorname{DPP}(\mathcal{K}_J) \biggl[ \prod_{i\in J} \lambda_i(L)\biggr] / \det{(L + I_n)}.
\]
The mixture weight of $\operatorname{DPP}(\mathcal{K}_J)$ is given by the product of the eigenvalues $\lambda_i(L)$ corresponding to the eigenvectors $v_i\in V_J$, normalized by $\det\left(L + I_n\right) = \prod_{i=1}^{n}\left[\lambda_i(L)+1\right]$.
Sampling of a subset $\boldsymbol{Y}\sim \operatorname{DPP}(L)$ can be realized by first selecting an elementary DPP, $\operatorname{DPP}(\mathcal{K}_J)$, with probability equal to its mixture component weight, and then, in a second step,
sampling a subset from $\operatorname{DPP}(\mathcal{K}_J)$. Moreover, it can be shown that the expected value and variance
of the number of elements in $\boldsymbol{Y}$, $\operatorname{card}(\boldsymbol{Y})$, are given by
\begin{equation*}
  \mathbb{E}\left[\operatorname{card}(\boldsymbol{Y})\right] = \sum_{i=1}^{n}\tfrac{\lambda_i(L)}{\lambda_i(L)+1} \; ; \;
  \operatorname{Var}\left[\operatorname{card}(\boldsymbol{Y})\right] =  \sum_{i=1}^{n}\tfrac{\lambda_i(L)}{(\lambda_i(L)+1)^2}.
\end{equation*}
It is well known that the computational complexity of obtaining the eigendecomposition of a $n\times n$ symmetric matrix is of order $O(n^3)$ and, as $n$ grows larger, the computation of the eigenvalues and eigenvectors
becomes expensive.
Even the storage of the matrix is limited by the memory required.\\

The sampling algorithm based on DPP starts with a subset of the eigenvectors of the kernel matrix, selected at random, where the probability of selecting each eigenvector depends on its associated eigenvalue. This fact suggests directly that it is unnecessary to compute all the eigenvalues, as eigenvectors with low associated eigenvalues are selected with low probability. This is particularly useful in the case of large matrices: computing only the largest eigenvalues can substantially reduce the computational burden of obtaining all the eigenvalues. The literature points to many references of well-known algorithms that can extract the $t$ largest (or smallest) eigenvalues, with their associated eigenvectors, of a $n \times n$ Hermitian matrix, where usually, we have $t\ll n$. One of the most classical and used algorithms is the Lanczos algorithm \citep{Lanczos1950}. Despite its popularity and computational efficiency, the algorithm was proven to be numerically instable, due in part to the loss of orthogonality of the computed Krylov subspace basis vectors generated. Since then, many efforts have been made to solve this issue, like \cite{Cullum1978}, \cite{Parlett1989} or \cite{Grimes1994}. Many of the variations of the Lanczos algorithm propose a restart after a certain number of iterations. One of the most popular restarted variations of the algorithm is the implicitly restarted Lanczos method, proposed by \cite{Calvetti1994}, which is implemented in \textsc{arpack} \citep{Lehoucq1998}, motivating then our preference for this particular variation.

The Lanczos algorithm and its implicitly restarted variation were specially developed for large sparse symmetric matrices. Consequently, our first step before implementing the implicitly restarted Lanczos method is to find a good approximation of the kernel matrix $L$ by a sparse matrix. We decide to address this question by following two approaches: one approach based on the Nearest Neighbor Gaussian Process \cite{Datta2016} and another approach based on random sampling of small submatrices from the dense kernel matrix $L$.


\subsection{The Nearest Neighbor Gaussian Process}\label{NNGP}

The Nearest Neighbor Gaussian Process (NNGP) was developed by \cite{Datta2016} and extended by \cite{Finley2019}, to obtain a sparse approximation of the kernel matrix $L$, say $\widetilde{L}$, to which the implicitly restarted Lanczos method can be applied to obtain its largest eigenvalues.

\cite{Finley2019} showed that the covariance matrix $W$ of a Gaussian process can be expressed through a specific Cholesky decomposition:
\begin{equation}\label{eq:finley}
  W =\left(I_n- A\right)^{-1} D\left(I_n-A\right)^{-T},
\end{equation}
where $A$ is a $n\times n$ strictly lower-triangular matrix and $D$ is a $n\times n$ diagonal matrix;
here $(\cdot)^{-T}$ stands for the inverse of the transposed matrix.
In order to define properly the matrices $A$ and $D$ we need to introduce the following notation.
For any $n\times n$ matrix $M$, and a subset of indices $J\subseteq \{1,\ldots, n\}$, we will write
$M_{k,J} = (M_{kj})_{j\in J}$ for the $\operatorname{card}(J)$-dimensional vector formed by the corresponding
components of the row $k$ of $M$, $k=1,\ldots, n$. We define similarly, $M_{J,k}$. Also, we write $[i_1:i_2]$ for the set
$J=\{ j : i_1 \leq j \leq i_2\}$.

Having introduced the notation, we can write the $i$th row of $A$, $A_{i\sbullet}$  as
$A_{i, [1:i-1]} = W_{[1:i-1]}^{-1} \, {W}_{[1:i-1], i},$ for $i=2,\ldots, n,$ and
$A_{i, [i:n]} = 0$, for $i=1,\ldots, n$.
where ${W}_{[1:k]}$ represents the leading principal submatrix of order $k$ of the matrix ${W}$. The diagonal entries $D_{ii}$ of ${D}$ are such that $D_{11}=W_{11}$ and
$D_{ii} = W_{ii} - W_{i,[1:i-1]} \, A_{i, [1:i-1]}^T$, for $i=2,\dots, n$.

Note that these equations are the linear equations that define the matrices $A$ and $D$. These equations need to be solve
for $A$ and $D$ in order to obtain the decomposition given by the expression in~\eqref{eq:finley}.
Unfortunately, the computation of ${A}_{i\sbullet}$ still takes $O(n^3)$ floating point operations,
specially for high values of $i$ closer to $n$, which increases the dimension of ${W}_{i-1}$.
Despite this shortcoming, the authors mention that this specific decomposition highlights where the sparseness can be exploited:
setting to zero some elements in the lower triangular part of ${A}$.
This is achieved by limiting the number of nonzero elements in each row of ${A}$ to a maximum of $m$ elements.

Let $N_i\subseteq \{1, \ldots, n\}$  be the set of indices $j < i$ for which $A_{i,j}\neq 0$. We constraint $N_i$ to
 have at most $m$ indices.
In this latter case, all elements of the $i$th row ${A_{i\sbullet}}$ of ${A}$ are zero, except for the elements
$A_{i,N_i} = W_{N_i}^{-1} \, W_{N_i, i}$, for $i=2,\ldots, n$, where ${W}_{N_i}$ is the principal submatrix of ${W}$ whose rows and columns are indexed by $N_i$.
For the diagonal entries, we have $D_{11}=W_{11}$ and $D_{ii}=W_{ii}-{W}_{i, N_i} \, {A}_{i, N_i}^T$, for $i=2,\dots, n$.

These latter equations form a linear system of size at most $m\times m$, with $m=\underset{i}{\max}\left(\operatorname{card}(N_i)\right)$. This new system can be solved for $A$ and $D$ in $O(n m^3)$ floating points operations.
Using these solutions gives rise to the sparse approximation to the precision matrix $W^{-1}$
\begin{equation}\label{precision}
  \widetilde{W}^{-1}=\left({I}_n-{A}\right)^{T}{D}^{-1}\left({I}_n-{A}\right).
\end{equation}
The inverse of $\widetilde{W}^{-1}$ is an approximation to $W$.
\cite{Datta2016} show in the context of spatial Gaussian processes, that $\widetilde{W}^{-1}$ has at most $n m(m+1)/2$ nonzero
entries. Thus, $\widetilde{W}^{-1}$ is sparse provided that $m \ll n$.
We can apply this result to develop an efficient determinantal consensus clustering (see Section~\ref{sec:determinantal:consensus}) when the data size $n$ is large.
Recall that the kernel matrix $L$ is a real symmetric positive semidefinite matrix.
When $L$ is also positive definite, it can be seen as a covariance matrix.
This is what we assume from now on. Therefore, it
can be approximated using the NNGP approach by a matrix $\widetilde{L}$ whose inverse $\widetilde{L}^{-1}$ is sparse.

For each $i\in\{2, \ldots, n\}$ consider the distances $d_{ij} = \lVert x_i - x_j\rVert$ for $j< i$.
Let $d_{i(1)} \leq d_{i(2)} \leq  \cdots \leq d_{i(i-1)}$ be the corresponding sequence of ordered distances.
In our model we set $N_i = \{ j : j< i,\, d_{ij} \leq d_{i(m)}\}$, for $i > m$; and set $N_i = \{1,\ldots, i-1\}$ for
$i\leq m$.
Let $\check{L}$ be the $m$-nearest-neighbor matrix whose row entries are given by $\check{L}_{ii} = L_{ii}$,
$\check{L}_{ij} = L_{ij}$ if $j\in N_i$, and $\check{L}_{ij} = 0$, otherwise.
We would like to stress here that the matrix $\widetilde{L}$ based on the neighborhoods $\{N_i\}_{i=1}^n$
is not the same as the $m$-nearest-neighbor matrix $\check{L}$.
We have adopted  $\widetilde{L}$ instead of $\check{L}$ for several reasons.
First, $\widetilde{L}$ is a dense approximation of $L$, whose inverse $\widetilde{L}^{-1}$ is sparse.
Second, $\widetilde{L}^{-1}$ is a sparse matrix with $O(nm^2)$ nonzero entries.
On the other hand, $\check{L}$ is a sparse matrix with  $O(n m_{nn})$ nonzero entries,
where $m_{nn}$ is the number of nearest neighbors.
Therefore, for a fixed level of sparseness, building $\widetilde{L}^{-1}$ requires many fewer nearest neighbors $m = O( \sqrt{m_{nn}})$
than building $\check{L}$.
Finally, in Section~\ref{experiments2}, we compute both the Frobenius distances \citep{Horn20123}
$\lVert L -\widetilde{L}\rVert_F$, and $\lVert L -\check{L}\rVert_F$, and show that the former
is always smaller than the latter,
for all our simulated data.
Moreover, in terms of the symmetrized Kullback-Leibler divergence \citep{Kullback1951}, the distribution of the eigenvalues of $\widetilde{L}$
is closer to the distribution of the eigenvalues of $L$ than the distribution of the eigenvalues of $\check{L}$.

As our primary goal is to obtain the $t \ll n$ largest eigenvalues of the kernel matrix $L$,
we start with the construction of the sparse matrix $\widetilde{L}^{-1}$  using the formula in~\eqref{precision},
and the sparse computation form above.
We then apply the Lanczos algorithm to extract the $t$ smallest eigenvalues of $\widetilde{L}^{-1}$, and
their associated eigenvectors.
By inverting the eigenvalues, we obtain the $t$ largest eigenvalues of $\widetilde{L}$; the eigenvectors of $\widetilde{L}$
are the same as those of $\widetilde{L}^{-1}$.
With the eigenvalues and eigenvectors in hand, we can proceed
as usual with the determinantal consensus clustering of Section \ref{sec:determinantal:consensus}.
We stress here that the actual DPP used for the determinantal consensus clustering after this construction
is $DPP_\mathcal{S}(\widetilde{L})$, and not $DPP_\mathcal{S}(L)$. In pratice, $L$ is chosen only to give us
a measure of similarity between data points.

\subsection{Approach based on random sampling of small submatrices from $L$}\label{knn}

In this section, we consider another approach to  deal with large datasets and kernel matrices.
In this approach we combine dimension reduction techniques and the advantages of working with sparse matrices.
Let $L^{(1)}, \dots, L^{(M)}$ denote $M$ $r\times r$ submatrices sampled uniformly at random
and without replacement from $L$, where $r<n$ (ideally, $r\ll n$).
The idea is to use these submatrices as proxies for $L$ in the DPP sampling.

By generating a sufficiently large number $M$ of matrices, we expect to cover the set of eigenvectors and eigenvalues
of $L$ with the smaller sets of eigenvectors and eigenvalues of the matrices collection $\{ L^{(i)} : i=1,\ldots, M\}$.
This might be the case if the data are well separated so that, although $L$ might be a {\it dense} matrix, its
hidden structure might be {\it sparse.} That is, many entries in $L$ might be small.
We could also achieve sparseness by thresholding the elements of $L$. However, the following approach
yielded better results in our experiments (not shown here), and hence, it is the second approach adopted
(the first having been described in the previous section).

We apply the following iterative methodology to the set of submatrices, for a number $N$ of times.
For $k=1,\ldots, N$:
\begin{enumerate}
\item Select an index $i_k$ from $\{1, 2, \dots, M\}$ at random (with replacement),
  and consider the submatrix $L^{(i_k)}$.

\item Build a sparse approximation $\widehat{L}^{(i_k)}$ of the submatrix $L^{(i_k)}$ by considering the $k$-nearest neighbors
  of each point associated with the rows of the submatrix; that is,
  $\widehat{L}^{(i_k)}_{ij} = L^{(i_k)}_{ij}$ if $x_j$ is one of the $k$-nearest-neighbors of $x_i$ or if $x_i$ is one of the
  the $k$-nearest-neighbors of $x_j$;  $\widehat{L}^{(i_k)}_{ij} =0$, otherwise.

 \item  Generate a subset sample ${Y}_{i_k}$ from a $\operatorname{DPP}(\widehat{L}^{(i_k)})$
   based only on the $t$ largest eigenvalues extracted with the Lanczos algorithm.

 \item Find the Voronoi cells of the $n$ data points based on the sampled $Y_{i_k}$ center points.
   This generates the $k$th partition of the algorithm.

\end{enumerate}
At the end of this procedure, we
apply the determinantal consensus clustering of Section~\ref{sec:determinantal:consensus}
to the set of $N$ partitions obtained.

The number $M$ of submatrices to be sampled must be chosen so that we get benefits from using the submatrices to sample the generator sets through DPP rather using the whole kernel matrix $L$. We know that the computational complexity of obtaining the eigendecomposition of the $n\times n$ kernel matrix is $O(n^3)$ operations.
On the other hand, the eigendecompositions of the $M$ $r\times r$ submatrices requires $O(M r^3)$ operations.
To obtain computational gains from the sampled submatrices, we must guarantee that
$Mr^3<n^3$, that is, $M<(r/n)^{-3}$. The quantity $\gamma = r/n$
is the proportion of points considered in the submatrices.
Since we would like to take full advantage of both the dimension reduction and speed,
we should work with values of $M\ll\gamma^{-3}$.
In our experiments, we set
$M=\lfloor\gamma^{-3}/2\rfloor$, where $\lfloor x\rfloor$ stands for the floor function.

\section{Experiments with large datasets}\label{experiments2}

  In this section, we apply both approaches presented in Section~\ref{largedatasets} to moderately large datasets
  in order to compare their results.
  The comparison and evaluation of the results are based on simulations.
  Because the data size depends on the nature of the problem,
  there is no clear definition of what is considered a ``large dataset''.
  Here, we consider two size values as large sizes: $n\in \{1000, 10000\}$.
  These values were chosen so as to obtain results that can be applied to moderetaly large real datasets,
  and  to be able to explore the effect on the results of using sparse kernel matrices instead of the original
  kernel matrices for determinantal consensus clustering.
  Moreover, the chosen data sizes keep the computation time within reasonable elapsed times for our computer resources.
  Our choices for data sizes do not necessarily constitute what is known as \emph{big data},
  a term popularized by \cite{Mashey1997} to describe datasets with sizes that go
  beyond the ability of commonly used software tools to capture, curate, manage, and process data within a tolerable elapsed time \citep{Snijders2012}.
  As \cite{Bonner2017} emphasize, the processing of large datasets does not have to involve big data.

\subsection{Large datasets with $n=1000$ observations}

\paragraph{Data generation.}\;
Following \cite{vicente&murua1-2020}, we created nine experimental conditions or scenarios to generate datasets with $n=1000$ observations. Each dataset was generated with the algorithm of \cite{Melnykov20123}, which draws datasets from $p$-variate finite Gaussian mixtures. These have the form
$\sum_{k=1}^K \pi_k \phi_p(\cdot; \mu_k, V_k)$, where $K$ is the number of Gaussian components,  $\phi_p$ denotes the $p$-variate Gaussian density, $\{{\mu}_1, \ldots, {\mu}_K\}$, are the component means, and
 $\{V_1, \ldots, V_K\}$ are the covariance matrices of the components.
The means constitute $K$ independent realizations of a uniform $p$-variate distribution on the $p$-dimensional unit hypercube;
the covariance matrices are  $K$ independent realizations of a $p$-variate standard Wishart distribution with $p + 1$ degrees of freedom; the mixing proportions $\pi_k$ are drawn from a Dirichlet
distribution, so that $\sum_{k=1}^{K} \pi_k =1$.
The number of data points per component is a draw from the multinomial distribution based on the mixing proportions.

\cite{Melnykov20123} introduce the concept of \emph{pairwise overlap} to generate datasets with the algorithm.
It represents the degree of interaction between components and defines the clustering complexity of datasets.
\cite{Melnykov20123} define a range of 0.001 to 0.4 for an average pairwise overlap between components, representing a very low and a very high overlap degree, respectively.
We follow the values suggested by the authors, and choose, for every dataset generated,  a random average pairwise overlap between 0.001 and 0.01.
This range of values keep the overlap degree at a low level so that performing clustering on the data makes sense.
All datasets were generated from mixtures with ellipsoidal covariance matrices; the simulated components do not necessarily
contain the same number of elements.

  To obtain the nine simulated datasets with $n=1000$ observations,
  we consider $p\in\{5, 10, 18\}$ variables, and $K\in\{4, 8, 15\}$ components.
  These values correspond to three different levels (low, medium, large) for $p$ and $K$.
We also ensured that no cluster with size less than $\sqrt{n}$ is present among each simulated dataset,
because otherwise, following the procedure described in Section~\ref{sec:consensus},
small clusters will be inevitably merged with larger clusters.
We applied the two approaches presented in Section~\ref{largedatasets} to each simulated dataset.

\paragraph{Results with approach based on NNGP}\;
In order to study the effect of sparseness in the
approximation presented in Section~\ref{NNGP},
we set different values for $m$, the maximum number of nonzero elements in each row of the matrix $A$. Each value ensures four levels of sparseness
(or total percentage of zeros) for the matrix $\widetilde{L}^{-1}$: $20\%, 40\%, 60\%$ and $80\%$.
The first largest $t$  eigenvalues and corresponding eigenvectors of $\widetilde{L}^{-1}$ were extracted with the Lanczos algorithm, for $t\in\{10, 25, 50\}$.
The consensus DPP of Section~\ref{sec:determinantal:consensus} was applied to obtain 200 partitions, as recommended in \citep{vicente&murua1-2020}.
The whole procedure was repeated ten times.
Considering the nine simulated scenarios,
the experiment consists of nine plots with ten repeated measures from a  $4\times 3$ factorial design given
by the combination of sparseness and number of extracted eigenvalues.
For each single plot, we note that all $120$ observations are dependent, since the same data were used for all
combinations.

\begin{figure}[h]
\centering
\subfloat[Sparseness=20\% ; $t$=10 eigenvalues]{%
\includegraphics[width=0.4\textwidth]{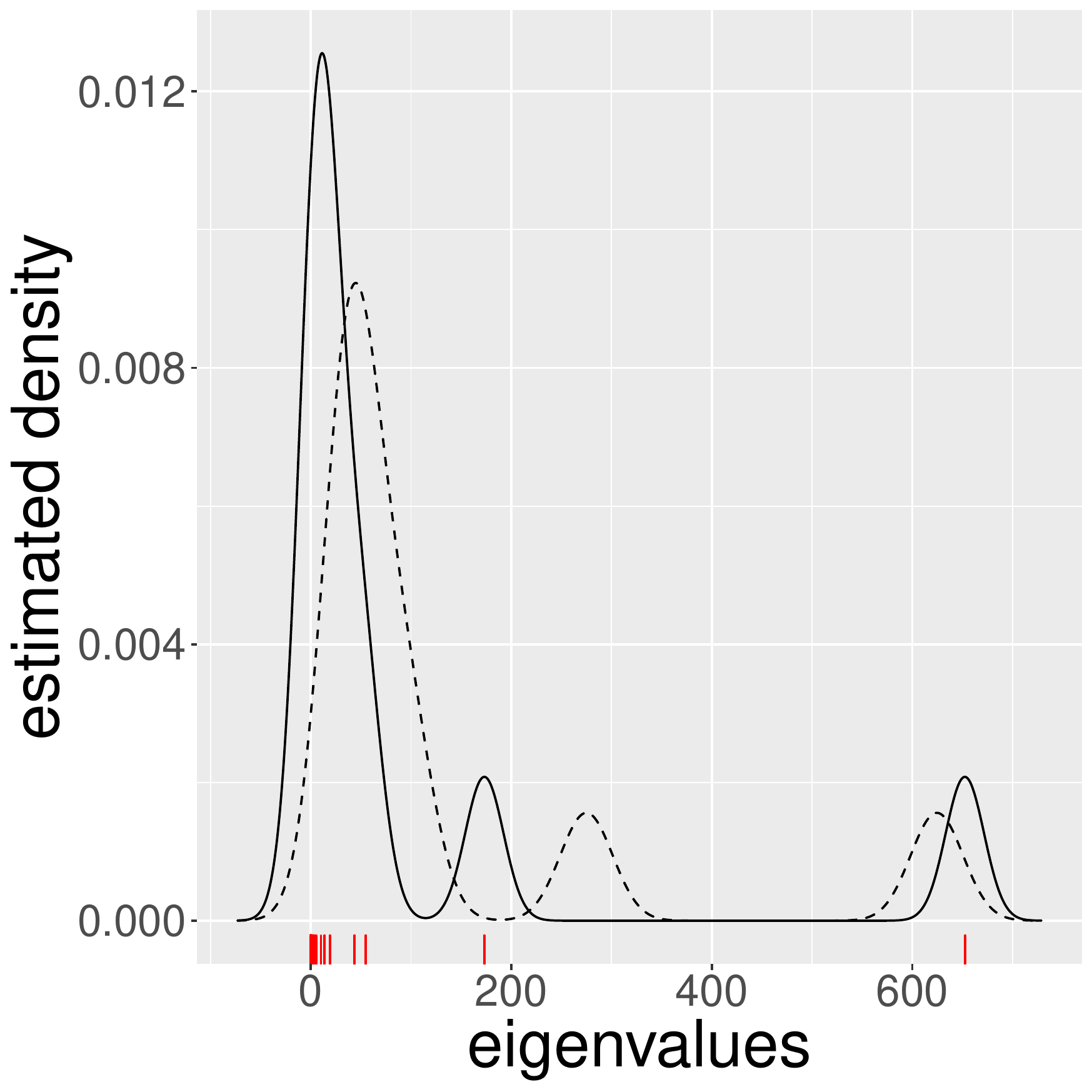}}
\quad
\subfloat[Sparseness=60\% ; $t$=50 eigenvalues]{%
\includegraphics[width=0.4\textwidth]{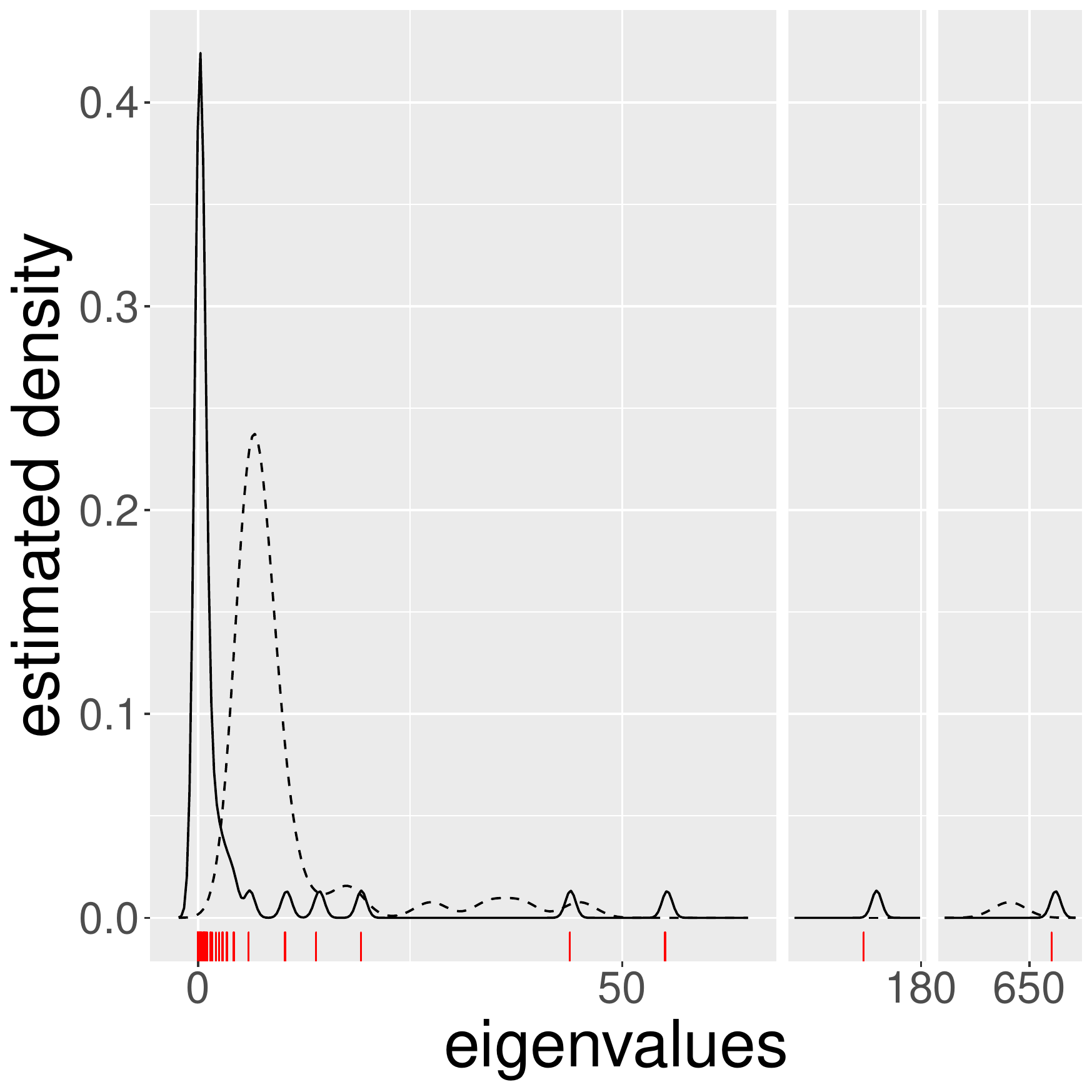}}
\caption{Kernel density estimates
  of the eigenvalue distribution from the NNGP approximation $\widetilde{L}$ (solid line), and the $m_{nn}$-nearest-neighbor matrix $\check{L}$ (dashed line). The plots are associated with two sparseness-eigenvalue  conditions among the
  twelve experimental conditions for a given dataset. The tick-marks indicate the eigenvalues of $L$. }
\label{densityestimNNGP}
\end{figure}

  Figure~\ref{densityestimNNGP} shows density estimates of the eigenvalues distribution obtained with the NNGP approximation matrix
  $\widetilde{L}$ (solid line).
  We considered two combinations of sparseness levels and number of eigenvalues: ($20\%, t=10$) and ($60\%, t=50$),
respectively.
The figure presents plots from one dataset, since they depict typical plots obtained with all datasets. The tick-marks in the horizontal axis locate all eigenvalues of the original kernel matrix $L$.
We can see that the density estimations obtained from $\widetilde{L}$ concentrate correctly around the true eigenvalues of the kernel matrix $L$.
For comparison purposes, we also display in Figure~\ref{densityestimNNGP} the density estimates of the set of eigenvalues from the
$m_{nn}$-nearest-neighbor matrix $\check{L}$ (dashed line) described in Section~\ref{NNGP}.
Unlike the NNGP approach, the eigenvalue density estimations associated with $\check{L}$ do not quite concentrate
on the set of true eigenvalues.
The plots corroborate the arguments of Section~\ref{NNGP} concerning our preference for $\widetilde{L}$ instead of $\check{L}$.

Following the discussion of Section~\ref{largedatasets},
we also computed the Frobenius distances between the sparse approximation matrices, $\widetilde{L}$ and $\check{L}$,
and the original dense matrix $L$.
Table~\ref{FDNNGP} shows the mean and standard deviation of Frobenius distances considering all nine data scenarios,
as a function of the level of sparseness.
The mean distance $\lVert L - \widetilde{L}\rVert_F$ is always much inferior to the distance $\lVert L - \check{L}\rVert_F$.
At first, the distances $\lVert L - \widetilde{L}\rVert_F$ decrease with sparseness until a level of at least $60\%$ sparseness is reached. However,
the distances of $L$ to $\check{L}$ monotonically increase with sparseness.
The approximation based on NNGP is then the preferred choice for extracting eigenvalues.
These results corroborate the nice concentration of the eigenvalue distributions of the NNGP approach
about the eigenvalues of $L$ in Figure~\ref{densityestimNNGP}.

\begin{table}[H]
      \centering

    \begin{tabular}{ccccc}
  \toprule
  \multirow{2}{*}{\begin{tabular}[c]{@{}c@{}}Approximation\\ matrix\end{tabular}}   & \multicolumn{4}{ c }{Sparseness level} \\\cmidrule{2-5}
   & $20\%$ & $40\%$ & $60\%$ & $80\%$\\ \midrule
  $\widetilde{L}$  & 11.31 (16.49) & 6.80 (4.55) & 8.04 (5.70) & 66.58 (77.14)\\
  $\check{L}$ & 200.17 (17.41) & 315.89 (16.27) & 421.82 (8.66) & 527.88 (8.18)\\ \bottomrule
\end{tabular}
    \caption{Mean and standard deviation (within parentheses) of Frobenius distances.    }
\label{FDNNGP}
\end{table}

To have an objective measure of the resemblance between the two sets of eigenvalues from $\widetilde{L}$ and $L$,
we computed the symmetrized Kulback-Leibler (KL) divergence \cite{Kullback1951} between corresponding density estimates of the two eigenvalue
distributions. The densities were estimated using a Gaussian kernel density estimator  \citep{Silverman1986}.
Table~\ref{KLNNGP} reports the mean and standard deviation of the KL divergences
 over all nine data scenarios
for each combination of sparseness and number of eigenvalues extracted.
There does not appear to be any significant difference between all the cases,
except for $t=10$ eigenvalues. Better results are obtained
when a larger but still very moderate number of eigenvalues is estimated.
The divergences reported in Table~\ref{KLNNGP} are very small,
showing that there is pretty good similarity between the eigenvalue distributions.
\begin{table}[H]
      \centering

    \begin{tabular}{ccccc}
  \toprule
  Number of & \multicolumn{4}{ c }{Sparseness level} \\\cmidrule{2-5}
  eigenvalues & $20\%$ & $40\%$ & $60\%$ & $80\%$\\ \midrule
  10 & 0.01 (0.00) & 0.36 (0.98) & 0.36 (0.97) & 0.35 (0.96)\\

  25 & 0.01 (0.00) & 0.01 (0.00) & 0.01 (0.00)  & 0.01 (0.00)\\

  50 & 0.01 (0.00) & 0.01 (0.00) & 0.01 (0.00)  & 0.01 (0.00)\\ \bottomrule
\end{tabular}
\caption{Mean and standard deviation (within parentheses) of Kullback-Leibler divergences.}
\label{KLNNGP}
\end{table}
We also report and compare the elapsed time in seconds for calculating each set of eigenvalues.
Table~\ref{timeNNGP} displays the means and standard deviations of elapsed times over all nine data scenarios
for each combination of sparseness and number of eigenvalues extracted.
We can see that, as expected, the average computation time decreases with sparseness, and increases with the number of eigenvalues extracted.
In fact, a linear regression (not shown here) of the elapsed time as a function of sparseness and number of eigenvalues extracted yields a coefficient of determination of 0.97,
clearly indicating
a linear growth of the computational time with both sparseness and the number of eigenvalues to be estimated.
The same statistics computed from the original kernel matrix $L$, from which all eigenvalues must be extracted, yield a mean of 0.27 seconds with a standard deviation of 0.06. Extracting only a few eigenvalues with the Lanczos algorithm reduces
significantly the computation time.
The elapsed times were computed while running a Julia v1.1.1 script \citep{Bezanson2017}
on a PC with an Intel-Core i5-4460 CPU running at 3.20GHz with 16GB of RAM.
\begin{table}[H]
\centering
\begin{tabular}{ccccc}
  \toprule
  Number of & \multicolumn{4}{ c }{Sparseness level} \\\cmidrule{2-5}
  eigenvalues & $20\%$ & $40\%$ & $60\%$ & $80\%$\\ \midrule
  10 & 0.10 (0.00) & 0.07 (0.00) & 0.06 (0.01) & 0.04 (0.01)\\

  25 & 0.12 (0.01) & 0.09 (0.00) & 0.08 (0.01)  & 0.05 (0.01)\\

  50 & 0.14 (0.01) & 0.12 (0.01) & 0.11 (0.01)  & 0.07 (0.01)\\ \bottomrule
\end{tabular}
\caption{Means and standard deviations (within parentheses) of elapsed times in seconds for eigenvalues calculation.}
\label{timeNNGP}
\end{table}

To obtain the optimal clustering configuration for each dataset, we used
the determinantal consensus clustering described in Section \ref{sec:determinantal:consensus}, meeting the minimal cluster size criterion of $\sqrt{n}$ and using the kernel-based validation index of \cite{Fan20103}.
The quality of the chosen optimal clustering configuration has been assessed with the adjusted Rand index or ARI
\citep{Rand19713,Hubert19853}.
Among the many known measures of goodness-of-fit that can be found in the literature,
the ARI is one of the most common criteria.
The original Rand index counts the proportion of elements that are either in the same clusters in both clustering configurations or in different clusters in both configurations.
The adjusted version of the Rand index corrected the calculus of the proportion, so that its expected value is zero when the clustering configurations are random.
The larger the ARI, the more similar the two configurations are, with the maximum ARI score of 1.0 indicating a perfect match.
Table~\ref{ressparseNNGP} displays the ARI means and standard deviations over all nine scenarios and replicas for
 all twelve combinations
of sparseness and number of eigenvalues extracted.

\begin{table}[H]
\centering
\begin{tabular}{cccc}
\toprule
\begin{tabular}[c]{@{}c@{}}Sparseness\\ level\end{tabular}  & $t=10$ & $t=25$ & $t=50$ \\ \midrule
20\%       & 0.94\,(0.06)   & 0.95\,(0.06)     & 0.95\,(0.06)     \\
40\%       & 0.93\,(0.05)   & 0.95\,(0.05)     & 0.95\,(0.06)    \\
60\%       & 0.94\,(0.05)   & 0.95\,(0.06)     & 0.95\,(0.05)     \\
80\%       & 0.93\,(0.06)   & 0.94\,(0.05)     & 0.95\,(0.05)
\\ \bottomrule
\end{tabular}
\caption{Global ARI means and standard deviations (within parentheses) yielded by consensus DPP with the sparse kernel matrices for the twelve combinations of sparseness and number of eigenvalues extracted. Each mean and standard deviation was computed from ninety datasets.}
\label{ressparseNNGP}
\end{table}

  For comparison purposes, we computed the ARI statistics obtained by applying
  consensus DPP with the original dense kernel matrix $L$, from which we extracted all the eigenvalues and eigenvectors.
  This yielded an ARI mean of 0.91, and a standard deviation of 0.08.
  We did the same with the consensus clustering methodology applied to partitions generated with the well-known
  Partitioning Around Medoids (PAM) algorithm \citep{Kaufmann19873}.
 The PAM algorithm is a classical partitioning technique for clustering.
 It chooses the data center points of the Voronoi cells by uniform random sampling.
 Because DPP selects center points based on diversity,
 our goal here is to show how much the quality of the clustering configurations is affected
 by the lack of diversity at the moment of  sampling centroids.
 PAM yielded a much lower ARI mean of 0.86, and a standard deviation of 0.14.

   We can see that the quality of the clustering results associated with the NNGP approach
   is better in terms of ARI than the ones yielded using the whole dense kernel matrix $L$.
   This conclusion holds regardless of the sparseness level and number of eigenvalues extracted.
   The results are also more stable, considering the standard deviations.
   Taking advantage of the elapsed time, choosing $t=25$ eigenvalues is the better choice among the three levels for $t$ studied here. This combined with a higher sparseness level (to speed up the computation of the matrix $\widetilde{L}^{-1}$)
appears to be a winning combination.
The results yielded by PAM are not as good. 
In conclusion, the NNGP approach provides  a very good and efficient alternative to the use of the complete dense matrix $L$.

\begin{figure}[H]
\centering
\subfloat[$\gamma=0.05$; Sparseness=20\% ; $t$=10 eigenvalues]{%
\includegraphics[width=0.3\textwidth]{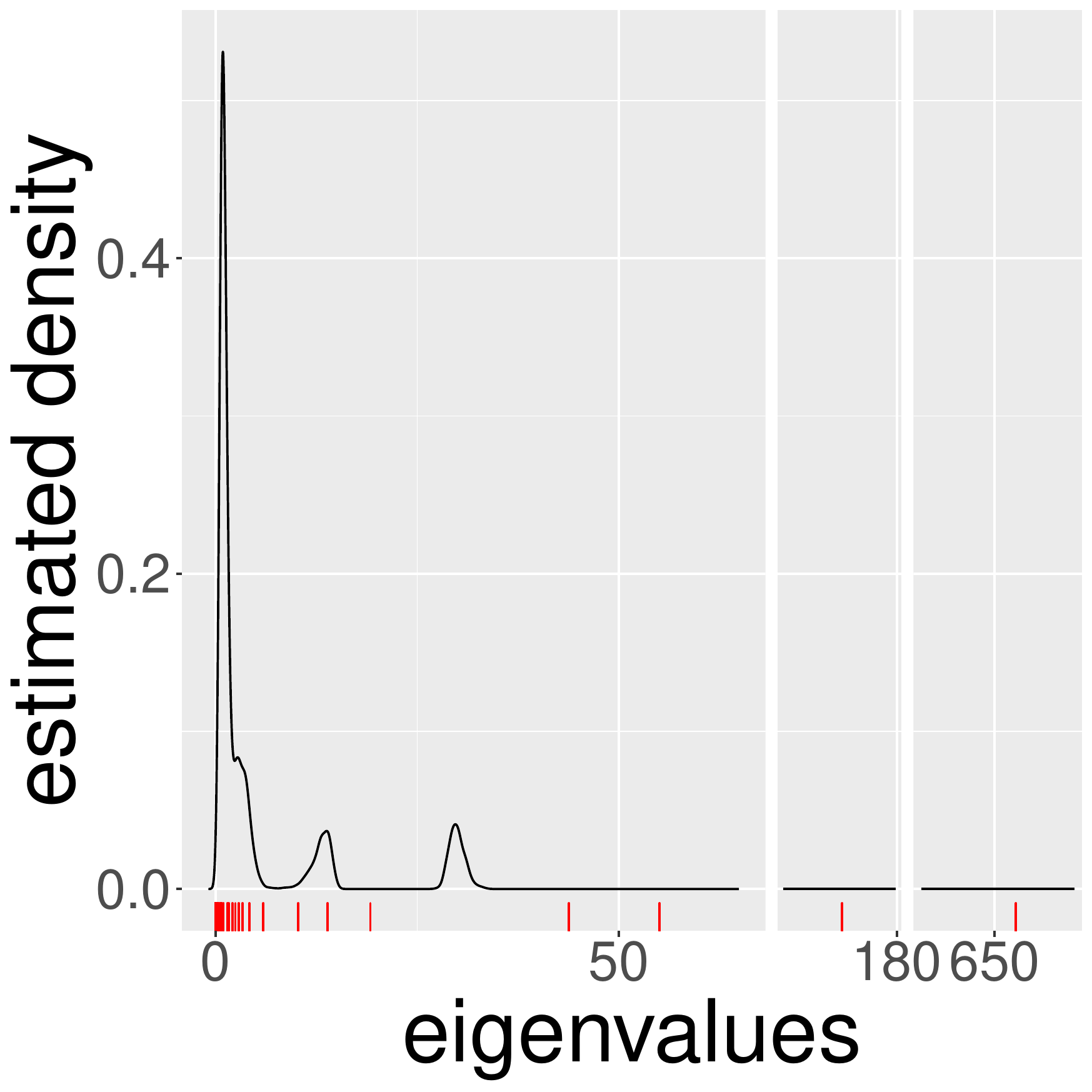}}
\quad
\subfloat[$\gamma=0.1$; Sparseness=40\% ; $t$=25 eigenvalues]{%
\includegraphics[width=0.3\textwidth]{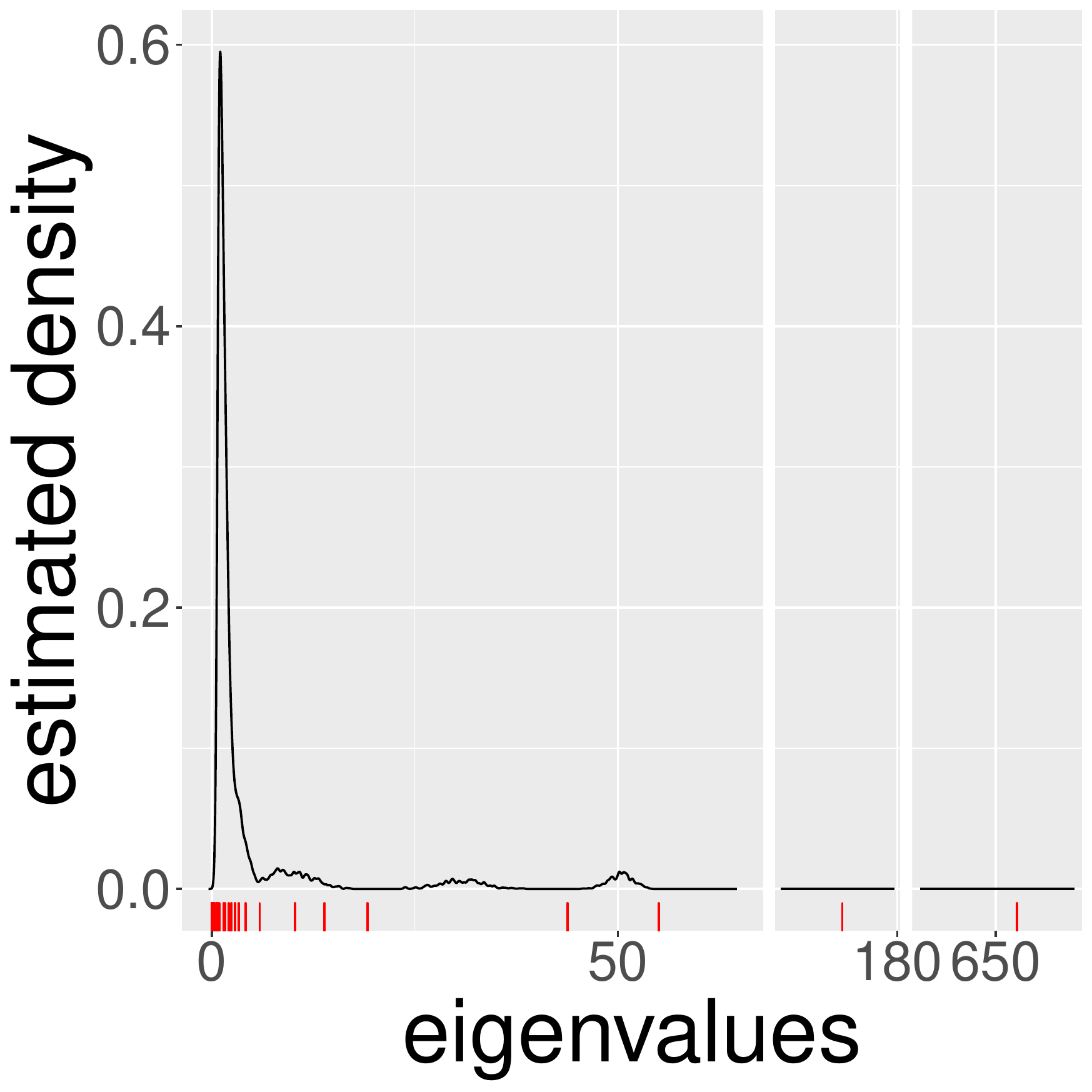}}
\quad
\subfloat[$\gamma=0.2$; Sparseness=60\% ; $t$=50 eigenvalues]{%
\includegraphics[width=0.3\textwidth]{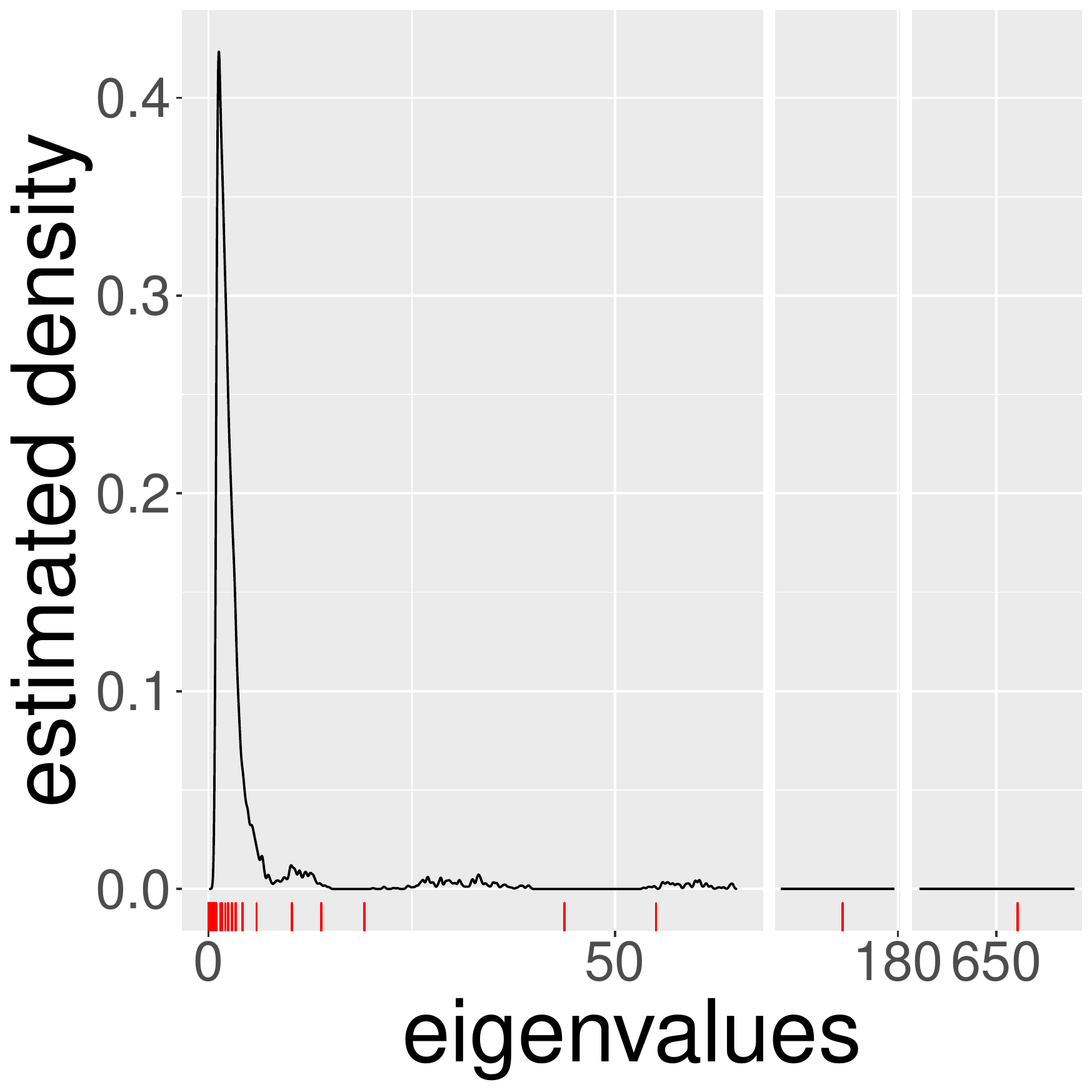}}
\caption{Density estimators of three sets of eigenvalues extracted from the random small matrices approach associated with three scenarios of Table~\ref{tabknn} on a given dataset. The tick-marks are placed on the eigenvalues of $L$.
}
\label{densityestimsmallGram}
\end{figure}

\paragraph{Results with approach based on small submatrices from $L$}\;
Following the notation presented in Section~\ref{knn}, we studied the effect of three parameters on the clustering results. These are the proportion $\gamma$ of points chosen (or, equivalently, the size $r$ of any submatrix $L^{(i_k)}, \; k=1, 2,\dots, N$), the sparsennes level of any submatrix $L^{(i_k)}$ (along with the number $k$ of nearest neighbors needed to achieve such level), and the number $t\in\{10, 25, 50\}$ of the largest eigenvalues to be extracted. Table~\ref{tabknn} shows the choices for $\gamma$ and sparseness levels with the corresponding values of $r$ and $k$.
Note that
each value of $k$ ensures the same sparseness for all nine data scenarios.
\begin{table}[H]
\centering
\begin{tabular}{ccccccc}
  \toprule
          &     & \mbox{\ } & \multicolumn{4}{c}{$k$} \\
$\gamma$  & $r$ &  & 20\% & 40\% & 60\% & 80\% \\ \midrule
0.05      & 50  &  & 34   & 25   & 17   & 8    \\
0.1       & 100 &  & 68   & 50   & 34   & 16   \\
0.2       & 200 &  & 136   & 100 & 68   & 32   \\ \bottomrule
\end{tabular}
\caption{Number of nearest neighbors $k$ associated with choices for proportion of data $\gamma$ and sparseness levels
  20\%, 40\%, 60\% and 80\%. }
\label{tabknn}
\end{table}
As we did with the NNGP approach,
we set the number of  partitions to obtain a consensus clustering to 200.
The whole procedure was repeated ten times.
The analysis of the results mimics the one made in the previous section with the NNGP approach.
Figure~\ref{densityestimsmallGram} shows density estimates of the eigenvalue distribution
associated with three combinations of
($\gamma$, sparseness, eigenvalues) $\in \{ (0.05, 20\%, 10),$ $(0.1, 40\%, 25),$ $(0.2, 60\%, 50)\}$.
  The plots are associated with a given dataset; it depicts typical patterns oberved in all datasets.
The tick-marks in the horizontal axis locate all eigenvalues of the original kernel matrix $L$.
We can see that the density estimations concentrate well around the smaller eigenvalues of $L$,
but fail to capture the largest eigenvalues.

 Table~\ref{KLknn} shows the Kullback-Leibler symmetrized divergences between the
 distribution of the eigenvalues extracted from the random small matrices and the
 distribution of the eigenvalues of $L$, for the ($\gamma,$ sparseness) combinations displayed in Table~\ref{tabknn}.
 The eigenvalues distributions were estimated with kernel density estimators.
 We can see that, globally, increasing the proportion $\gamma$ of points sampled reduces the KL divergence
 when combined with moderate to low levels of sparseness and a higher number of eigenvalues.
 Overall, the KL divergences are larger than those obtained with the NNGP approach.
 But, again, we stress that the objective of the approaches is not to estimate the eigenvalues of $L$,
 but to offer efficient DPP sample alternatives to DPP$_\mathcal{S}(L)$.
 \begin{table}[H]
\centering
\begin{tabular}{cccccc}
\toprule
\multirow{2}{*}{$\gamma$} & \multirow{2}{*}{\begin{tabular}[c]{@{}c@{}}Number of\\ eigenvalues\end{tabular}} & \multicolumn{4}{c}{Sparseness level} \\ \cmidrule{3-6}
 &  & 20\% & 40\% & 60\% & 80\% \\ \midrule
\multirow{3}{*}{0.05} & 10 &  2.09 (1.28) & 1.28 (0.37) & 1.00 (0.08) & 1.01 (0.03)  \\
 & 25 & 0.02 (0.01) & 0.58 (1.25) & 1.64 (0.46) & 1.01 (0.04) \\
 & 50 & 0.02 (0.01) & 0.03 (0.23) & 1.71 (0.88) & 1.10 (0.17)\\ \midrule
\multirow{3}{*}{0.1} & 10 & 1.95 (1.39) & 1.34 (0.43) & 1.02 (0.10) & 1.00 (0.03)  \\
 & 25 & 0.01 (0.00) & 0.12 (0.58) & 1.87 (0.89) & 1.04 (0.10)  \\
 & 50 & 0.02 (0.00) & 0.01 (0.00) & 0.15 (0.68) & 1.32 (0.26)\\ \midrule
\multirow{3}{*}{0.2} & 10 & 2.07 (1.43) & 1.36 (0.44) & 1.03 (0.12) &  1.00 (0.03)  \\
 & 25 & 0.01 (0.00) & 0.16 (0.69) & 1.76 (1.08) & 1.12 (0.23)  \\
 & 50 & 0.01 (0.00) & 0.01 (0.00) & 0.01 (0.00) & 1.64 (0.51)  \\ \bottomrule
\end{tabular}
\caption{Means and standard deviations (within parentheses) of Kullback-Leibler divergences associated with the random small matrices approach.}
\label{KLknn}
 \end{table}
Recalling the notation of Section \ref{knn}, Table~\ref{timeknn} shows the means and standard deviations of the
elapsed times in seconds for eigenvalues computation, using the sampled submatrix $L^{(i_k)}$ or its sparse approximation $\widehat{L}^{(i_k)}$.
 The results were obtained while running a Julia script on a PC with an Intel Core i5-4460 CPU running at 3.20GHz
 with 16GB of RAM.
 Note that contrary to NNGP, the level of sparseness does not affect the elapsed times for computing the eigenvalues
 with the Lanczos method. This fact is probably due to the small size of the matrices $\widehat{L}^{(i_k)}$.
 \begin{table}[H]
\centering
\begin{tabular}{ccccc}
\toprule
\multirow{2}{*}{$\gamma$} & \multirow{2}{*}{Submatrix} & \multicolumn{3}{c}{Number of eigenvalues} \\ \cmidrule{3-5}
 &  & 10 & 25 & 50  \\ \midrule
\multirow{2}{*}{0.05} & $L^{(i_k)}$ & 0.001 (0.000) & 0.001 (0.000) & 0.001 (0.000)  \\
 & $\widehat{L}^{(i_k)}$ & 0.001 (0.000) & 0.001 (0.000) & 0.001 (0.000)  \\ \midrule
\multirow{2}{*}{0.1} & $L^{(i_k)}$  & 0.004 (0.003) & 0.004 (0.003) & 0.004 (0.003)  \\
 & $\widehat{L}^{(i_k)}$ & 0.003 (0.001) & 0.003 (0.001) & 0.003 (0.001)\\ \midrule
\multirow{2}{*}{0.2} & $L^{(i_k)}$  & 0.01 (0.004) & 0.01 (0.004)  & 0.01 (0.004)  \\
 & $\widehat{L}^{(i_k)}$ & 0.006 (0.002) & 0.006 (0.002) & 0.006 (0.002)   \\ \bottomrule
\end{tabular}
\caption{Mean and standard deviation (within parentheses) of elapsed times in seconds to eigenvalues calculation, for $L^{(i_k)}$ and its sparse approximation $\widehat{L}^{(i_k)}$.
}
\label{timeknn}
 \end{table}

Table~\ref{ressparseknn} displays the ARI means and standard deviations over all
($\gamma$, sparseness) combinations in Table~\ref{tabknn}.
  For each combination, these statistics were computed from a sample of ninety ARI scores given
  by the ten replica datasets from the nine data scenarios.
  Table~\ref{resdense} displays the same statistics obtained by keeping the dense version of the submatrices $L^{(i_k)}$,
  and extracting all its eigenvalues, instead of using their sparse approximations.
  This comparison is appropriate to study the effect of making these small matrices sparse.
  These results should be compared to those of
  (i) consensus DPP applied to the original dense kernel matrix $L$, from which all eigenvalues are extracted,
  and (ii) the consensus clustering methodology applied to partitions generated by PAM.
  These latter results were already mentioned above when showing the results of the  NNGP approach.
  They are, respectively,  (i) a mean ARI and standard deviation of 0.91 and 0.08, and
  (ii)  a mean ARI and standard deviation of 0.86 and 0.14.

\begin{table}[H]
\centering
\begin{tabular}{ccccc}
\toprule
$\gamma$ & \begin{tabular}[c]{@{}c@{}}Sparseness\\ level\end{tabular} &  $t=10$ & $t=25$ & $t=50$ \\ \midrule
  \multirow{4}{*}{0.05}
 & 20\% &  0.92 (0.06) & 0.94 (0.05) &  0.96 (0.06)\\
 & 40\% &  0.93 (0.06) & 0.95 (0.05) &  0.95 (0.05)  \\
 & 60\% &  0.94 (0.05) & 0.95 (0.05) &  0.95 (0.06)  \\
 & 80\% &  0.93 (0.06) & 0.95 (0.06) &  0.95 (0.06) \\ \midrule
  \multirow{4}{*}{0.1}
  & 20\% &  0.93 (0.05) & 0.93 (0.06) &  0.94 (0.07)  \\
  & 40\% &  0.95 (0.06) & 0.95 (0.05) &  0.95 (0.06) \\
 & 60\%  &  0.95 (0.05) & 0.95 (0.05) &  0.95 (0.06) \\
 & 80\%  &  0.94 (0.06) & 0.94 (0.07) &  0.95 (0.06)  \\ \midrule
  \multirow{4}{*}{0.2}
  & 20\% &  0.93 (0.05) & 0.94 (0.06) &  0.93 (0.07)  \\
 & 40\%  &  0.94 (0.05) & 0.95 (0.05) &  0.94 (0.06)  \\
 & 60\%  &  0.94 (0.05) & 0.94 (0.05) &  0.92 (0.06)  \\
& 80\%   &  0.94 (0.06) & 0.95 (0.06) &  0.91 (0.08) \\
  \bottomrule
\end{tabular}
\caption{Global ARI means and standard deviations (within parentheses) associated with consensus DPP on the sparse random small submatrices, for each ($\gamma$, sparseness) combinations in Table~\ref{tabknn}, and over the corresponding
  ninety datasets for each combination.}
\label{ressparseknn}
\end{table}

\begin{table}[H]
\centering
\begin{tabular}{ccc}
\toprule
$\gamma=0.05$ & $\gamma=0.1$ & $\gamma=0.2$ \\ \midrule
0.94 (0.05) & 0.95 (0.05) & 0.95 (0.06) \\ \bottomrule
\end{tabular}
\caption{Global ARI means and standard deviations (within parentheses) associated with consensus DPP on the dense version of the random small submatrices considering all datasets.}
\label{resdense}
\end{table}

The quality of the clustering results is much better in terms of ARI, if we use either the dense $L^{(i_k)}$
or sparse $\widehat{L}^{(i_k)}$ random small submatrices rather than the whole dense kernel matrix $L$.
The sparse approximations $\widehat{L}^{(i_k)}$ require more computation since the nearest neighbors must be computed.
This extra cost does not seem worth when comparing the results associated with the dense approximations $L^{(i_k)}$.
However, there is a slight improvement in the results when $\gamma=0.05$.
In this case, combining any sparseness level with a higher number of eigenvalues is generally a good combination,
since computational times for eigenvalue extraction do not differ significantly.
For other levels of $\gamma$, the gain is not worth the complication of making the submatrices sparse.

Again, as with the  NNGP approach, the random small submatrices approach outperforms PAM and shows less variability in the quality of the results.
Furthermore, this approach achieves
comparable quality results to those of the NNGP approach.
From a computational point of view, the NNGP approach is more expensive because it requires dealing with larger matrices.
Therefore, the random small submatrices are a good and very efficient alternative to the use of the complete dense matrix $L$,
and to the use of the NNGP approach.

\subsection{Large dataset with $n=10000$ observations}

\paragraph{Data generation.}\;
For the second experiment, due to hardware limitations, we simulated only two datasets with $n=10000$ observations.
As in the previous case, these were generated with the algorithm of \cite{Melnykov20123}.
The first one has $p=15$ variables and $K=10$ components, while the second one has $p=10$ variables and $K=5$ components.
Throughout this section, these data will be referred to as dataset I and dataset II, respectively.
Both datasets have a maximum pairwise overlap of 0.01.
We ensured that no cluster with size less than $\sqrt{n}$ is present among the simulated datasets, as it will be inevitably merged with a larger cluster, according to the procedure described in Section~\ref{sec:consensus}. We applied the two approaches presented in Section~\ref{largedatasets} to the simulated datasets.

\paragraph{Results with approach based on NNGP}\label{sec:NNGP10000}\;
We studied the same four levels of sparseness for the matrix $\widetilde{L}^{-1}$ ($20\%, 40\%, 60\%, 80\%$), setting
the appropriate values for the maximum number $m$ of nonzero elements in each row of the matrix $A$ (Section~\ref{NNGP}).
The first largest $t\in\{100, 250, 500\}$ eigenvalues and corresponding eigenvectors of $\widetilde{L}^{-1}$ are extracted with the Lanczos algorithm. The determinantal consensus clustering of Section~\ref{sec:determinantal:consensus}
was then applied so as to obtain 200 partitions of the data.
The whole procedure was repeated five times.
This time, the experiment consists of two plots of five repeated measures each, from
 a $4\times 3$ factorial design given by the combination of four levels of sparseness and three
levels for the number of extracted eigenvalues.
All $60$ observations of each plot are dependent, since the same datasets were used for all scenarios.
The analysis of the results follows the same steps as the NNGP approach applied to the smaller datasets.

\begin{figure}[H]
\centering
\subfloat[Sparseness: 20\% ; $t$=100 eigenvalues]{%
\includegraphics[width=0.4\textwidth]{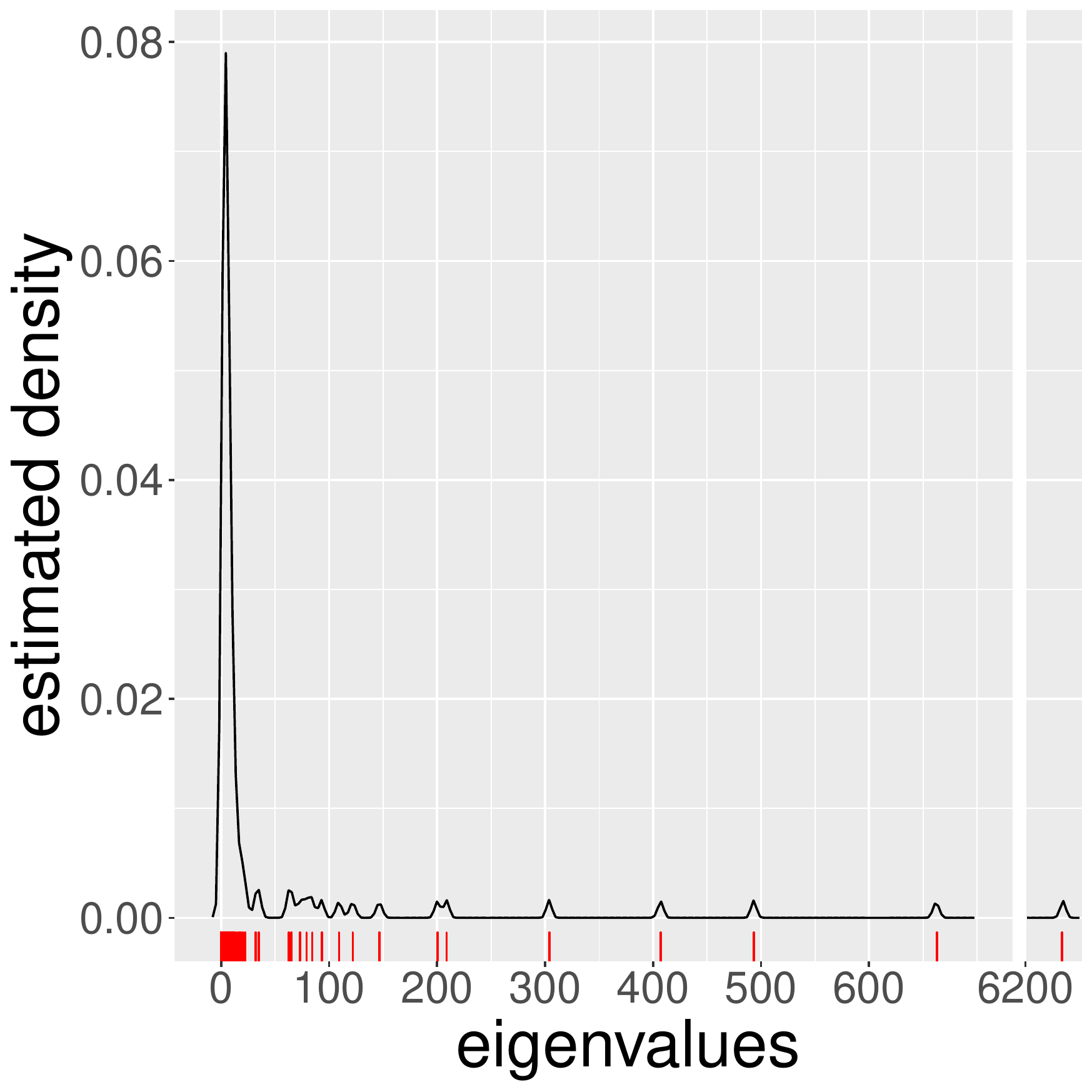}}
\quad
\subfloat[Sparseness: 60\% ; $t$=500 eigenvalues]{%
\includegraphics[width=0.4\textwidth]{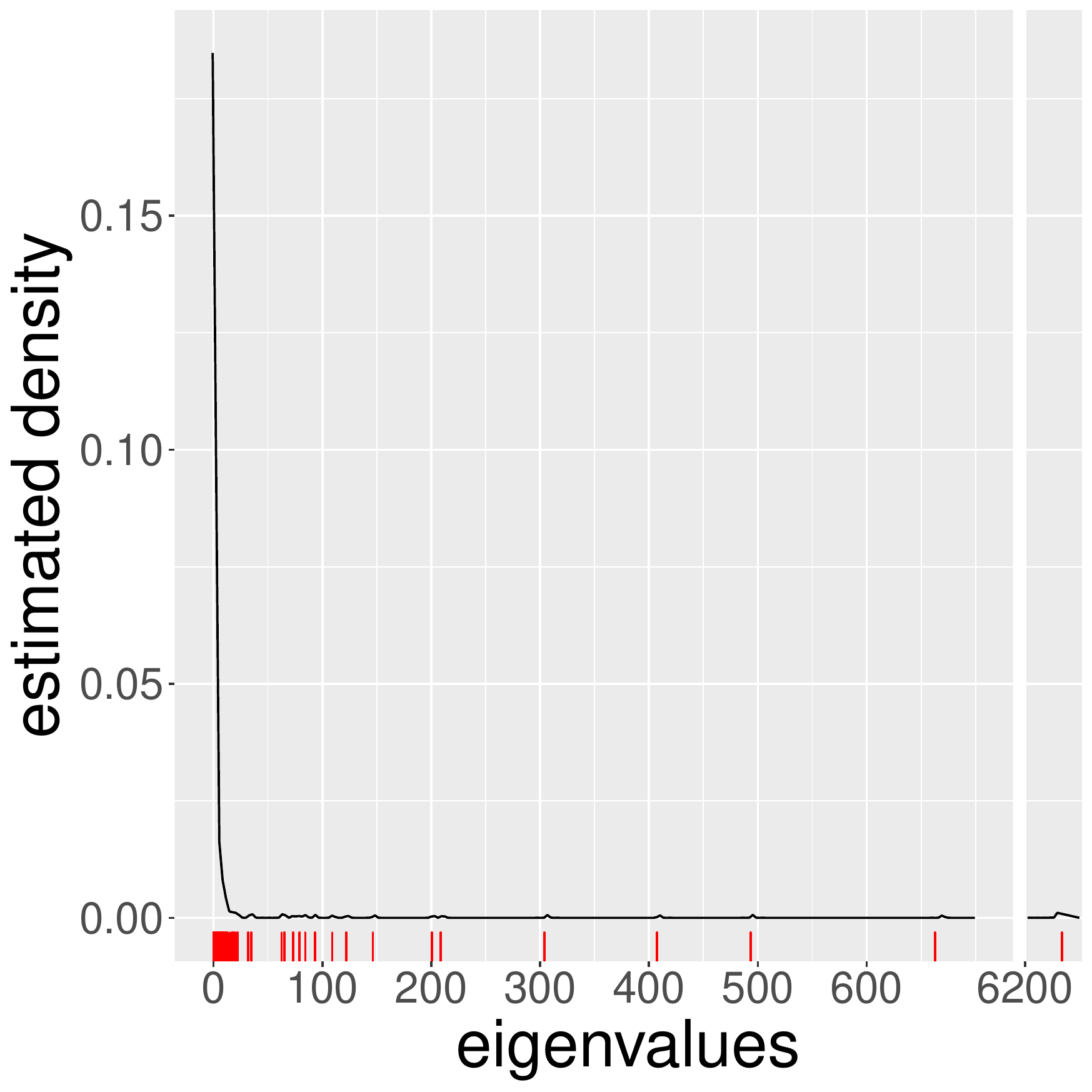}}
\caption{Kernel density estimates
  of the eigenvalue distribution associated with $\widetilde{L}$ for dataset I, for two sparseness-eigenvalue
  conditions among the twelve experimental conditions. The tick-marks indicate the eigenvalues of $L$.}
\label{densityestimNNGP10000}
\end{figure}

Figure~\ref{densityestimNNGP10000} shows density estimates of the eigenvalue distribution of $\widetilde{L}$ for dataset I,
for two combinations of sparseness and number of eigenvalues: ($20\%, t=100$) and ($60\%, t=500$),
respectively. We can see that the density estimations concentrate correctly around the true eigenvalues of the kernel matrix $L$.

For comparison purposes between the two sparse approximations  $\widetilde{L}$ and $\check{L}$,
we computed the Frobenius distance between these matrices and the kernel matrix $L$.
Table~\ref{FDNNGP10000} displays the results for both datasets. As seen in the experiment with the smaller datasets,
the distance of $L$ to $\widetilde{L}$ is always much smaller than the distance of $L$ to $\check{L}$.
For both datasets, the distances always increase with sparseness.
The approximation based on NNGP is a better choice for extracting eigenvalues.
This also explains the correct concentration of the density estimation of the eigenvalue distribution of $\widetilde{L}$
around the eigenvalues of $L$, as seen in Figure~\ref{densityestimNNGP10000}.

\begin{table}[H]
\centering
  \begin{tabular}{cccccc}
\toprule
\multirow{2}{*}{\begin{tabular}[c]{@{}c@{}}Approximation\\ matrix\end{tabular}} & \multirow{2}{*}{Dataset} & \multicolumn{4}{c}{Sparseness level} \\\cmidrule{3-6}
 &  & $20\%$ & $40\%$ & $60\%$ & $80\%$ \\ \midrule
\multirow{2}{*}{$\widetilde{L}$} & I &  2.75 & 8.82 & 15.65 & 96.71  \\
 & II & 0.33 &4.47  &14.07  &191.45  \\\cmidrule{3-6}
\multirow{2}{*}{$\check{L}$} & I & 2122.11 & 3269.61 & 4265.59 & 5224.48 \\
 & II &1955.09  &3079.10  &4128.15  &5228.46\\\bottomrule
\end{tabular}
\caption{Frobenius distances for both datasets (I and II).}
\label{FDNNGP10000}
\end{table}

Table~\ref{KLNNGP10000} reports the symmetrized Kullback-Leibler divergence
  of both datasets between the density estimates of the two
eigenvalue distributions from $\widetilde{L}$ and $L$. The densities were estimated with kernel density
estimators.
Again, as observed with the smaller datasets, the very small divergence values indicate a good resemblance
between the eigenvalue distributions.

\begin{table}[H]
\centering
  \begin{tabular}{cccccc}
\toprule
\multirow{2}{*}{\begin{tabular}[c]{@{}c@{}}Number of\\ eigenvalues\end{tabular}} & \multirow{2}{*}{Dataset} & \multicolumn{4}{c}{Sparseness level} \\\cmidrule{3-6}
 &  & $20\%$ & $40\%$ & $60\%$ & $80\%$ \\ \midrule
\multirow{2}{*}{100} & I &  0.0000338 & 0.0000337 &  0.0000337 &  0.0000336 \\
 & II & 0.0000329 & 0.0000132  & 0.0000015  & 0.0000329  \\\cmidrule{3-6}
\multirow{2}{*}{250} & I & 0.0000365 & 0.0000359 & 0.0000359  & 0.0000358 \\
 & II &0.0000133  &0.0000015  &0.0000329  &0.0000139\\\cmidrule{3-6}
 \multirow{2}{*}{500} & I & 0.0000090 & 0.0000090 & 0.0000090 & 0.0000090 \\
 & II &0.0000016  &0.0000322  &0.0000166  &0.0000017\\
 \bottomrule
\end{tabular}
\caption{Kullback-Leibler divergences for both datasets (I and II).}
\label{KLNNGP10000}
\end{table}

Table \ref{timeNNGP10000} shows the comparison of the elapsed times in seconds for calculating each set of eigenvalues
for dataset I.
Extracting all eigenvalues from the original matrix $L$ yields an elapsed time 170.08 seconds and 156.74 seconds for datasets I and II, respectively. The results were obtained on Julia v1.1.1 running on a PC with an Intel-Core i5-4460 CPU running at 3.20GHz with 16GB of RAM.
The elapsed times for dataset I are well explained by a linear regression on sparseness and number of eigenvalues extracted, presenting a coefficient of determination of $0.998$. Similar results were obtained for dataset II.
\begin{table}[H]
\centering
  \begin{tabular}{cccccc}
  \toprule
  \multirow{2}{*}{\begin{tabular}[c]{@{}c@{}}Number of\\ eigenvalues\end{tabular}}& \multirow{2}{*}{Dataset} & \multicolumn{4}{ c }{Sparseness level} \\\cmidrule{3-6}
 &  & $20\%$ & $40\%$ & $60\%$ & $80\%$\\ \midrule
  \multirow{1}{*}{100} & I &  37.34 &34.46 & 31.79 & 27.39\\
  \multirow{1}{*}{250} & I   & 68.03 & 65.86  & 60.62 &55.37 \\
  \multirow{1}{*}{500} & I   & 125.48 & 120.27  & 115.55 &112.85\\
  \\ \bottomrule
\end{tabular}
\caption{Elapsed times in seconds for eigenvalues calculation of dataset I. }
\label{timeNNGP10000}
\end{table}

\begin{table}[H]
\centering
\resizebox{\textwidth}{!}{\begin{tabular}{ccccccc}
\toprule
\multirow{2}{*}{\begin{tabular}[c]{@{}c@{}}Sparseness\\ level\end{tabular}} & \multicolumn{2}{c}{$t=100$} & \multicolumn{2}{c}{$t=250$} & \multicolumn{2}{c}{$t=500$} \\
 & I & II & I & \multicolumn{1}{c}{II} & I & \multicolumn{1}{c}{II} \\ \midrule
20\%                                                     & 0.97\,(0.04) & 0.79\,(0.04)  & 0.93\,(0.02)&  0.87\,(0.05)   & 0.70\,(0.04)& 0.88\,(0.07) \\
40\%                                                     & 0.98\,(0.00)& 0.87\,(0.03)  & 0.94\,(0.02)&  0.87\,(0.08)   & 0.77\,(0.03)& 0.89\,(0.04)   \\
60\%                                                     & 0.97\,(0.04) & 0.86\,(0.09) & 0.94\,(0.01)&  0.87\,(0.07)   & 0.83\,(0.02) &  0.82\,(0.07)  \\
80\%                                                     & 0.98\,(0.02)& 0.86\,(0.09)   & 0.97\,(0.02) & 0.86\,(0.04)     & 0.71\,(0.10) & 0.87\,(0.03)    \\ \bottomrule
\end{tabular}}
\caption{ARI means and standard deviations (within parentheses) obtained by
consensus DPP on the sparse kernel matrices over the twelve experimental conditions and both datasets (I and II).}
\label{ressparseNNGP10000}
\end{table}
Table~\ref{ressparseNNGP10000} displays the ARI means and standard deviations over all twelve experimental conditions
for both datasets (I and II).
For comparison purposes, we computed the same statistics for (i) consensus DPP
applied to the original dense kernel matrix $L$, from which all eigenvalues were extracted,
and for (ii) the consensus clustering methodology applied to partitions generated by PAM.
For dataset I, the corresponding ARI means and standard deviations are
0.57 and 0.02 for consensus DPP,
and 0.93 and 0.01 for PAM.
For dataset II, we obtain ARI means and standard deviations of
0.82 and 0.06
for consensus DPP, and of
0.83 and 0.03
for PAM.

Observe that the quality of the clustering results is better if we consider the sparse approximation $\widetilde{L}^{-1}$ and a lower number of eigenvalues, $t\in\{100, 250\}$. Considering a higher number of eigenvalues
is not such a good option. The quality of the results decreases with the number of eigenvalues extracted for each sparseness level. The use of the original dense matrix $L$ with all eigenvalues is not a good alternative either.
The most probably reason for the observed results is the large size of the kernel matrix.
Most eigenvalues cannot be computed reliably, numerically speaking;
hence, using all numerically extracted eigenvalues might introduce noise,
leading the algorithm to perform poorly.
Also, if most of the eigenvalues are small, they will be estimated with a lot of  error,
specially if the difference between the largest eigenvalues and the smallest ones is orders of magnitude.
This is known as ill-conditioning of the kernel matrix.
  We would like to note that for all three cases, NNGP, the original dense matrix and PAM,
  we used the $\sqrt{n}$ criterion to merge small clusters during consensus (see Section~\ref{sec:consensus}).
  However, as pointed out in \cite{vicente&murua1-2020}, a criterion close to $n^{2/3}$ might have been more appropriate
  given the small number of clusters $K$, and the large number of observations $n$.
  In fact, using this larger size
  of small cluster in the merging step of the consensus give slightly better results for all methods, specially for dataset II.

  To summarize, determinantal consensus clustering with the NNGP approach outperforms PAM for all cases.
  In addition, if we would like to favor high quality clustering results with low computational cost,
  combining $t=100$ eigenvalues with a high level of sparseness appears to be the best option.

\paragraph{Results with approach based on small submatrices from $L$}\;
For this approach, due to hardware limitations,
we decided to reduce the number of evaluated scenarios in both datasets.
Thus, maintaining the proportion $\gamma\in\{0.05, 0.1, 0.2\}$ of points selected from the dense matrix $L$,
the sampled submatrices $L^{(i_k)},\; k=1,\dots,N$,
are of size $r \in\{500, 1000, 2000\}$, respectively.
These sizes are considerably larger than those used with the smaller datasets.
Hence, sparse approximations of $L^{(i_k)}$ are used with a unique high level of sparseness equal to $80\%$.
This sparseness is achieved with 80, 160 and 320 nearest neighbors, respectively.
Recall that with the NNGP approach applied to the smaller datasets,
a low mean elapsed time for eigenvalue calculation was achieved when combining a low number of eigenvalues
with a high level of sparseness (see Table~\ref{timeNNGP}).
Therefore, to speed up the computation time, we extract only the largest 100
eigenvalues for all the cases.
The following analysis of the results is organized as with the above simulations.

\begin{figure}[H]
\centering
\subfloat[$\gamma=0.05$ ]{%
\includegraphics[width=0.3\textwidth]{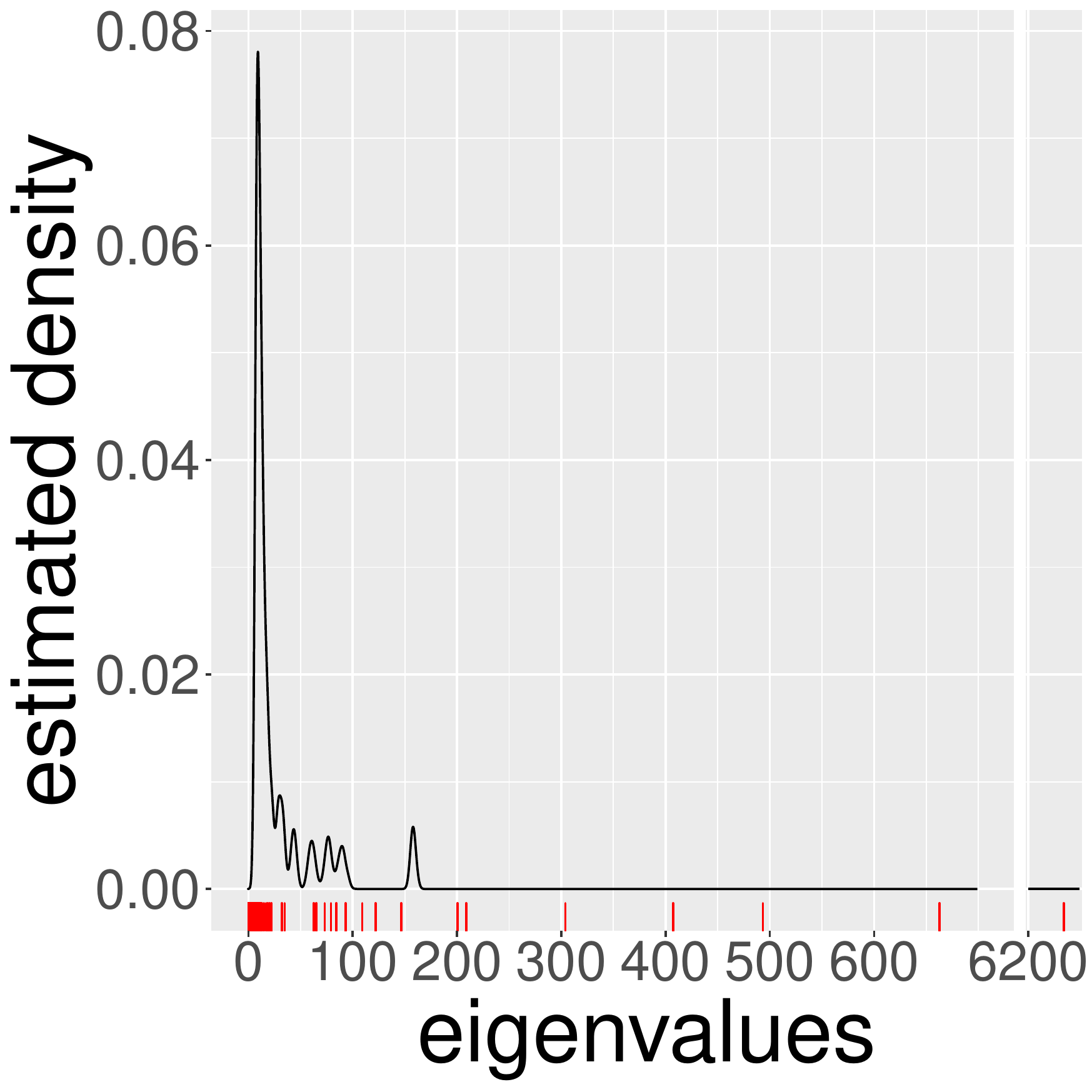}}
\quad
\subfloat[$\gamma=0.1$ ]{%
\includegraphics[width=0.3\textwidth]{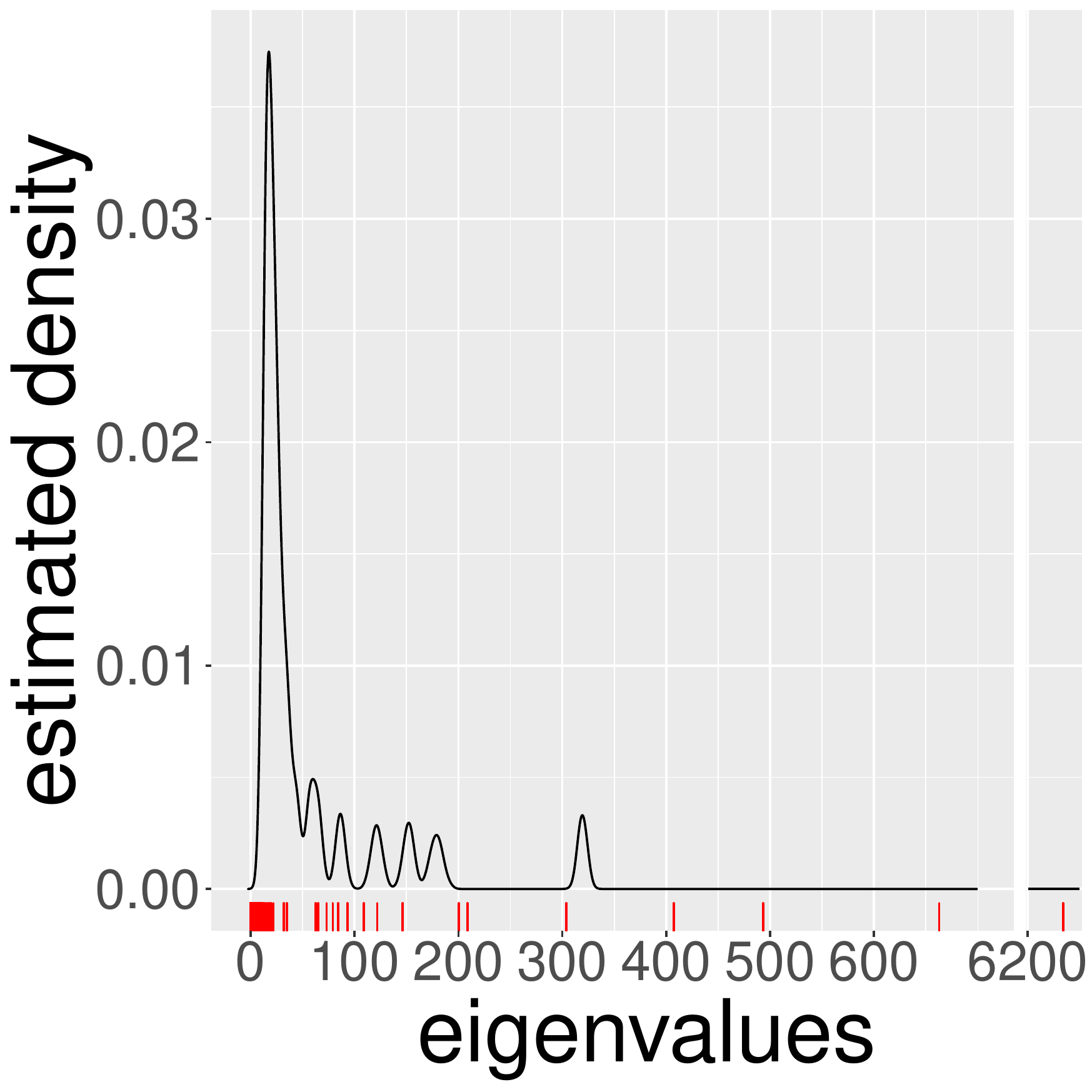}}
\quad
\subfloat[$\gamma=0.2$ ]{%
\includegraphics[width=0.3\textwidth]{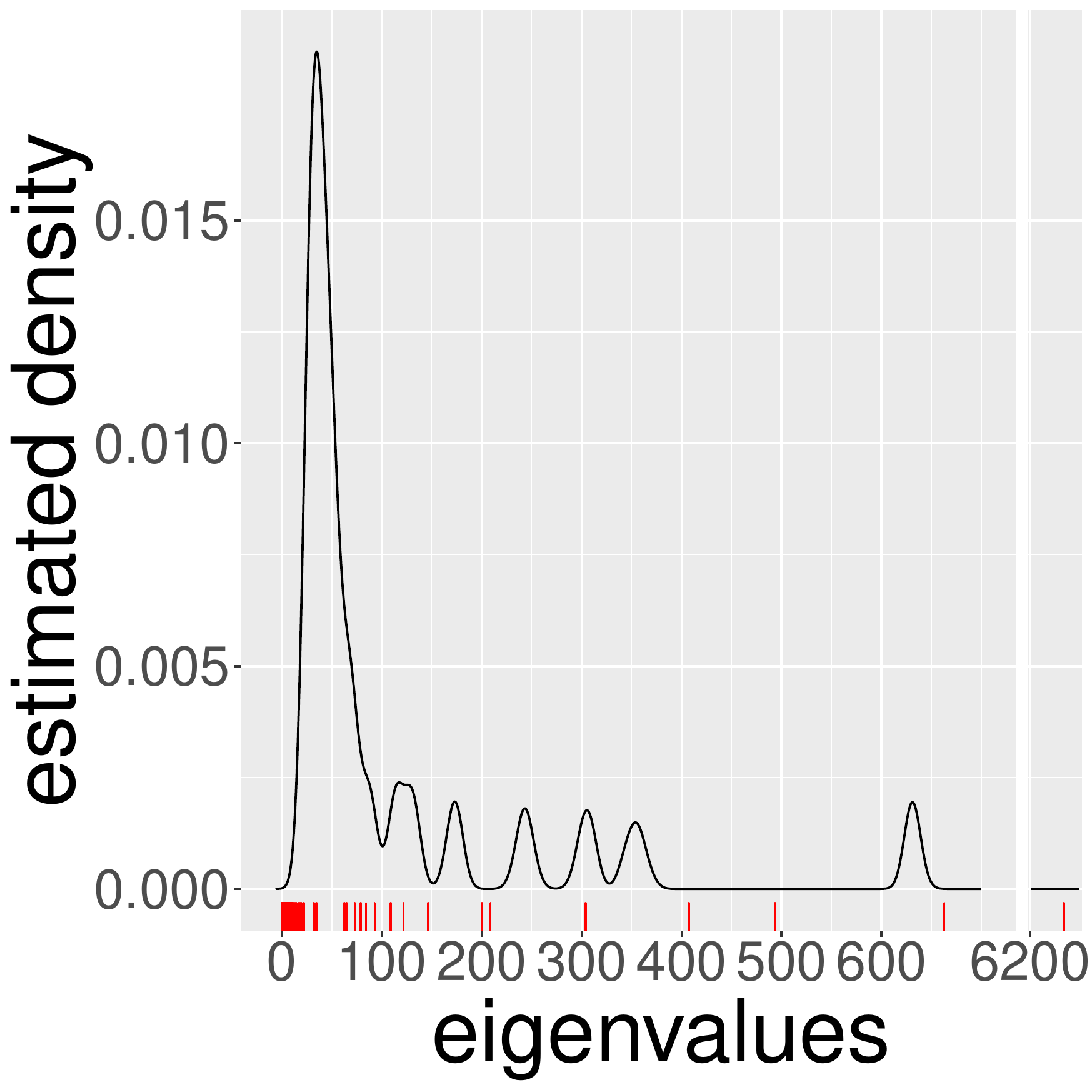}}
\caption{Kernel density estimates
of the set of eigenvalues extracted from sparse $L^{(i_k)}$. The tick-marks are placed on all eigenvalues of $L$ .}
\label{densityestimsmallGram10000}
\end{figure}

Figure~\ref{densityestimsmallGram10000} shows the kernel density estimates for the aforementioned situations. The tick-marks in the
horizontal axis locate all eigenvalues of the original kernel matrix $L$.
As seen in the results with the smaller datasets,
the density estimates of this approach do not capture well the largest eigenvalues when the sparseness is too high (for a low value of $\gamma$).
Table~\ref{KLknn10000} reports the symmetrized Kullback-Leibler divergences between the distribution
of the eigenvalues extracted from the sparse submatrices $L^{(i_k)}$ and
the distribution of the eigenvalues of $L$, for $\gamma\in\{0.05, 0.1, 0.2\}$. The densities were
estimated with kernel density estimators. The KL divergences decrease with $\gamma$ and are very small, indicating a good resemblance between the two distributions.
\begin{table}[H]
 \centering
 \begin{tabular}{cccc}
\toprule
Dataset & $\gamma=0.05$ & $\gamma=0.1$ & $\gamma=0.2$ \\ \midrule
I & 0.000875 & 0.000301 & 0.000095 \\
II & 0.000579 & 0.000173 & 0.000061\\
\bottomrule
\end{tabular}
\caption{Kullback-Leibler divergences for both datasets (I and II).}
\label{KLknn10000}
\end{table}

Table~\ref{timeknn10000} shows the comparison of the elapsed time in seconds for eigenvalue computation,
using the sampled submatrix $L^{(i_k)}$ or its sparse approximation $\widehat{L}^{(i_k)}$. The results were obtained with Julia Language, version 1.1.1, on a Desktop PC with Intel Core i5-4460 CPU @ 3.20GHz Processor and 16 GB DDR3 RAM.
In this case, and due to the relatively large size of the matrices $L^{(i_k)}$, it does make a difference to use the sparse matrices  $\widehat{L}^{(i_k)}$ instead of the dense ones.

\begin{table}[H]
\centering
  \begin{tabular}{ccccc}
\toprule
Dataset & Submatrix & $\gamma=0.05$ & $\gamma=0.1$ & $\gamma=0.2$ \\ \midrule
\multirow{2}{*}{I} &$L^{(i_k)}$  &  0.14 & 0.22 & 1.02  \\
 & $\widehat{L}^{(i_k)}$ & 0.03 & 0.08 & 0.30\\ \midrule
\multirow{2}{*}{II} & $L^{(i_k)}$ & 0.17 & 0.38 & 1.66 \\
 & $\widehat{L}^{(i_k)}$ & 0.09 &0.35  & 1.33 \\ \bottomrule
\end{tabular}
\caption{Elapsed times in seconds for eigenvalues calculation of datasets I and II. }
\label{timeknn10000}
\end{table}

  Table~\ref{ressparseknn10000} displays the ARI means and their standard deviations for $\gamma\in\{0.05, 0.1, 0.2\}$ and both datasets (I and II). It also reports the results obtained with consensus DPP applied to the original dense kernel matrix $L$, from which all eigenvalues were extracted,
and PAM. These latter results were already reported in Section~\ref{sec:NNGP10000}.
\begin{table}[H]
\centering
\begin{tabular}{cccccc}
\toprule
Dataset & $\gamma=0.05$ & $\gamma=0.1$ & $\gamma=0.2$ & Original $L$ & PAM \\ \midrule
I & 0.95 (0.00) & 0.96 (0.01) & 0.96 (0.01) & 0.57 (0.02) & 0.93 (0.01) \\
II & 0.88 (0.02) & 0.87 (0.04) & 0.87 (0.03) & 0.82 (0.06) & 0.83 (0.03)\\ \bottomrule
\end{tabular}
\caption{ARI means and standard deviations (within parentheses) obtained from consensus DPP with approach based on small submatrices from $L$ for datasets I and II. The results for the original dense matrix and PAM are also displayed. }
\label{ressparseknn10000}
\end{table}

As we did with the smaller datsets,
we show in Table~\ref{resdense10000}, the same statistics obtained by keeping the dense version of the submatrices $L^{(i_k)}$, and extracting all its eigenvalues, instead of using their sparse approximations.
\begin{table}[H]
\centering
\begin{tabular}{cccc}
\toprule
Dataset & $\gamma=0.05$ & $\gamma=0.1$ & $\gamma=0.2$ \\ \midrule
I & 0.94 (0.05) & 0.95 (0.05) & 0.95 (0.06) \\
II & 0.88 (0.04) & 0.81 (0.06) & 0.83 (0.05)\\  \bottomrule
\end{tabular}
\caption{Global ARI means and standard deviations (within parentheses) of consensus DPP on the dense version of the random small submatrices for datasets I and II.}
\label{resdense10000}
\end{table}

The random small matrices approach yields excellent results for any proportion $\gamma$.
If we consider the results with the dense version of the submatrices, we can see that the results obtained are similar or better.
As already observed with the small samples, the use of a sparse approximation $\widehat{L}^{(i_k)}$ does not hurt the quality of the results.
The main advantage of using the sparse approximations is the reduced amount of time needed to compute the eigenvalues.
Another advantage is the resulting stability of the clustering quality results (low dispersion of ARI values).
In summary, if one would like low computational time and low dispersion,
using the sparse approximation of submatrices with a low proportion of sampled points ($5\%$) is a good option.

To end this section, we observe that the approach based on random sampling of submatrices outperforms PAM, for all $\gamma$ values in both datasets.
It also reaches comparable
  quality clustering levels to the approach based on NNGP with $t=100$ and a sparseness of $80\%$.
  Sampling submatrices of lower dimension is then a good alternative to the use of the NNGP approach.

\subsection{DPP as a measure of diversity}

\begin{figure}[H]
\centering
\subfloat[Dataset with $n=1000$ observations]{%
\includegraphics[width=0.45\textwidth]{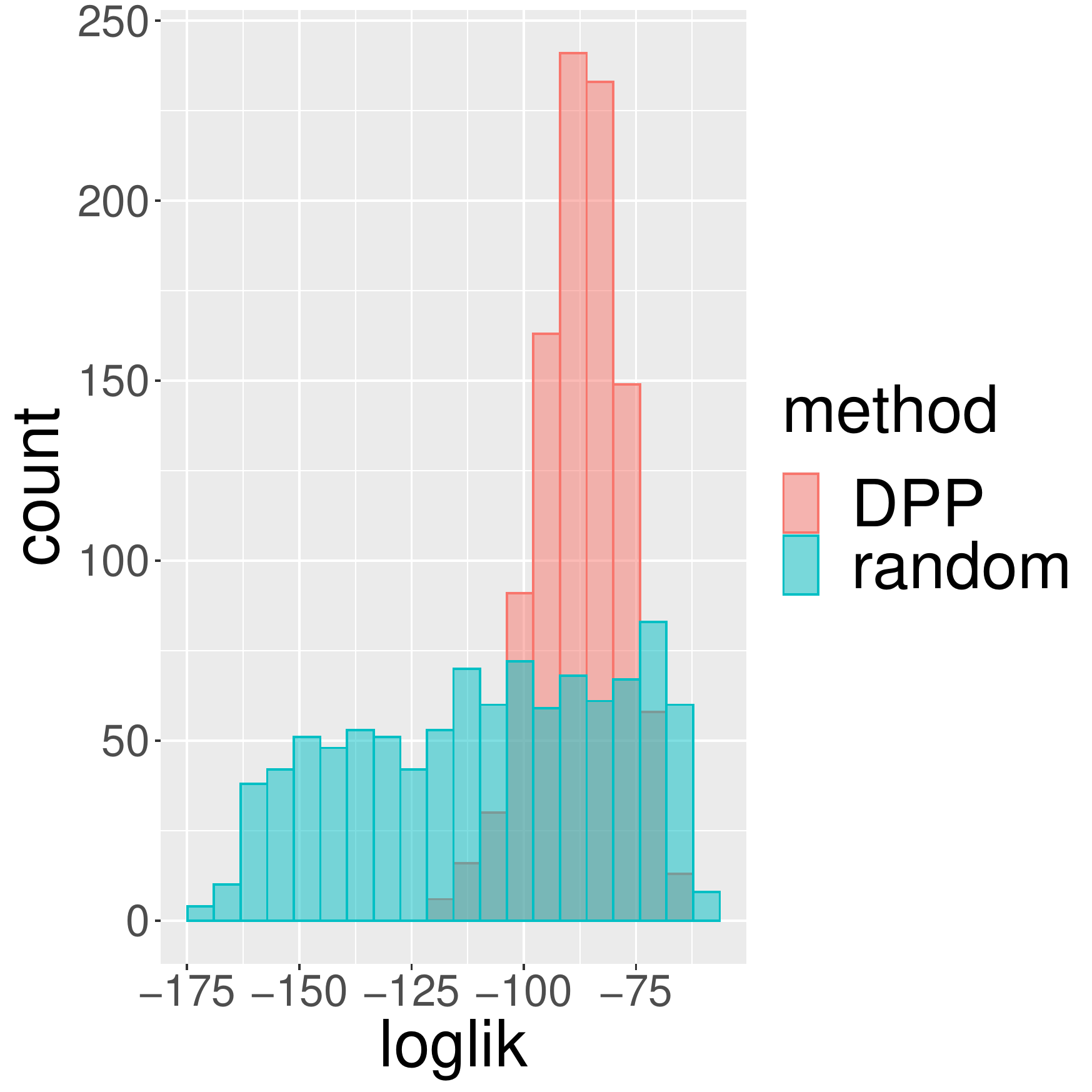}}
\quad
\subfloat[Dataset with $n=10000$ observations]{%
\includegraphics[width=0.45\textwidth]{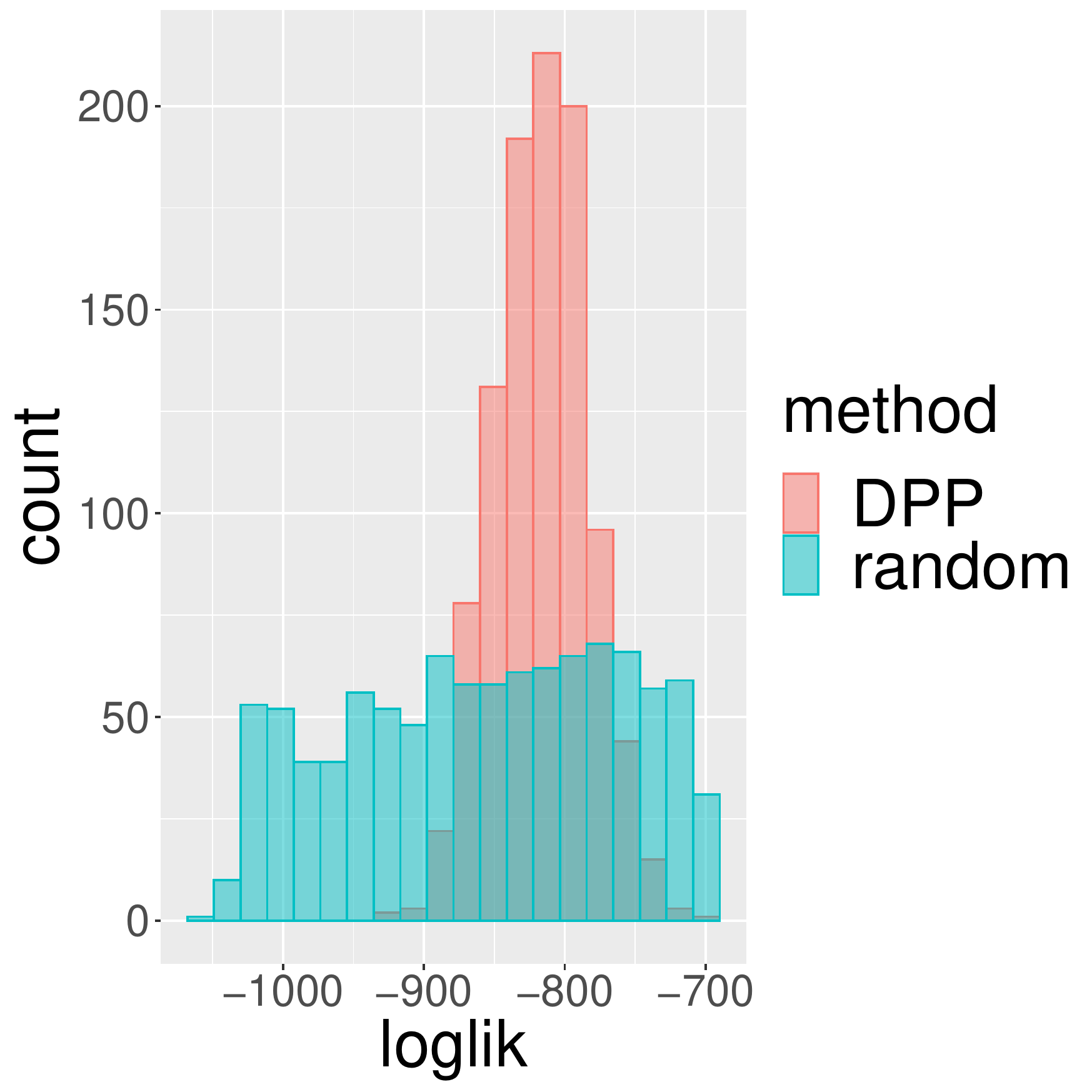}}
\caption{Histograms of the logarithm of the probability mass function (loglik), using DPP and simple random sampling, for two simulated large datasets.}
\label{diversitysim}
\end{figure}

The differences between DPP and PAM can also be highlighted using the logarithm of the probability mass function of the DPP, given by \eqref{definition2}. Figure~\ref{diversitysim} displays histograms of these probability logarithms.
The histograms are based on 1000 random subsets of datapoints drawn (i) using the DPP sampling algorithm of \cite{BenHough20063} and \cite{Kulesza20123}, and (ii) using the simple random sampling of PAM.
The subsets were drawn from two simulated large datasets: one of size $n=1000$, and another of size $n=10000$ observations.
The histograms clearly show that DPP selects random subsets with higher and less dispersed probability mass values (likelihood) than simple random sampling. This explains the low dispersion of the ARI observed in most results of this section, when sampling is performed with DPP. The higher likelihood achieved by DPP also implies a higher diversity of the sampled subsets. On its turn, subsets sampled as in PAM yield a highly dispersed likelihood, resulting in highly or poorly diverse subsets. DPP tends to be more consistent and stable since it ensures a high level of diversity among the selected elements at each sampling.

\section{Application to real data}\label{sec:real}

In this section we evaluate the performance of consensus DPP versus PAM on three large real datasets:
\vspace{3pt}

\noindent 1.- A dataset about human activity recognition and postural transitions using smartphones, collected from 30 subjects who performed six basic postures (downstairs, upstairs, walking, laying, sitting and standing), and six transitional postures between static postures (stand-to-sit, sit-to-stand, sit-to-lie, lie-to-sit, stand-to-lie and lie-to-stand). The experiment was realized in the same environment and conditions, while carrying a waist-mounted smartphone with embedded inertial sensors. The dataset consists of 10929 observations, with 561 time and frequency extracted features, which are commonly used in the field of human activity recognition.
  The dataset is available on the \emph{UCI Machine Learning Repository} \citep{Dua20193}, a well known database in the machine learning community for clustering and classification problems.
  The six transitional postures between static postures comprises a relatively small subset of observations.
  Therefore, we apply our clustering algorithm  only to the six basic postures.
  Using the notation of the simulated datasets of Section~\ref{experiments2}, the final dataset has $n=10411$ observations, $p=561$ variables and $K=6$ components.
  \vspace{3pt}

\noindent 2.- The Modified National Institute of Standards and Technology (\textsc{mnist}) dataset \citep{Lecun2010}, one of the most common datasets used for image classification. This dataset contains 60000 training images and 10000 testing images of handwritten digits, obtained from American Census Bureau employees and American high school students.
  Each 784-dimensional observation represents a $28\times 28$ pixel gray-scale image depicting a handwritten version of one of the ten possible digits (0 to 9).
  As it is a common practice with the  \textsc{mnist} dataset, due to its intrinsic characteristics,
  we transformed the data to its multiplicative inverse as hinted by the Box-Cox transformation.
  We only use the testing set of 10000 images, so that
  the final dataset has $n=10000$ observations, $p=784$ variables and $K=10$ components.
  \vspace{3pt}

\noindent 3.- The Fashion-\textsc{mnist} dataset \citep{Xiao2017}, also one of the most common datasets used for image classification. This dataset contains 60000 training images and 10000 testing images of Zalando's articles\footnote{%
Zalando is a European e-commerce company specializing in fashion.
They provided image data in repositories like Github.}.
  Each observation represents a $28\times 28$ pixel gray-scale images of clothes associated with a label from 10 classes.
  For the same reasons as \textsc{mnist}, we transformed the data with an appropriate Box-Cox transformation,
  and only worked with the testing portion of the data.
  The dataset has $n=10000$ observations, $p=784$ variables and $K=10$ components.
\vspace{6pt}

We applied both approaches of Section~\ref{largedatasets} to each dataset, and followed the same analysis procedure
of Section \ref{experiments2}. For the NNGP approach, we set the maximum number $m$ of nonzero elements in each row of the matrix $A$ to obtain a sparse approximation of the kernel matrix $L$ with $80\%$ of sparseness. The Lanczos algorithm was applied to extract the $t=100$ largest eigenvalues from the sparse approximated matrix.
For the approach based on random small submatrices from $L$, we sampled a proportion $\gamma=0.05$ of points from the kernel matrix $L$, and obtained sparse approximations of the sampled submatrices with $80\%$ of sparseness.
The Lanczos algorithm was applied to extract the $t=100$ largest eigenvalues from the sparse approximated submatrices.
Determinantal consensus clustering was applied to 200 partitions. The whole procedure was repeated five times.

For comparison purposes,
we also include the results from a couple of popular algorithms: consensus clusterings with PAM, and $k$-means.
The PAM algorithm was already mentioned in Section~\ref{experiments2}.
The $k$-means algorithm was introduced by Stuart Lloyd in 1957, with a supporting publication in \cite{Lloyd19823}.
Given an initial set of $k$ means, representing $k$ clusters,
it assigns each observation to the cluster with the nearest mean,
and proceeds to recompute means from observations in the same cluster.
The procedure is repeated until no changes are observed in the assignments.
Among the many algorithms to initialize the $k$ centers of $k$-means  \citep{Forgy19653,Pena19993,Gonzalez19853,Katsavounidis19943},
we chose the $k$-means$++$ algorithm of \cite{Arthur20073}.
This method has become largely popular \citep{Capo20173,Franti20193}.
It is a probability-based technique that avoids the usually poor results found by standard $k$-means initialization methods.
We performed consensus clusterings with PAM and $k$-means using 200 partitions, also  repeating the whole procedure five times.
We report the mean and the standard deviation of the ARI achieved by the different methods in Table~\ref{realdatasets}.

\begin{table}[H]
\centering
\begin{tabular}{ccccl}
\toprule
Dataset & NNGP & \begin{tabular}[c]{@{}c@{}}Small\\ submatrices\end{tabular} & PAM & $k$-means \\ \midrule
Smartphones & 0.59 (0.01) & 0.58 (0.01)  & 0.43 (0.12) &  0.49 (0.02)\\
\textsc{mnist} & 0.58 (0.05) & 0.36 (0.02) & 0.24 (0.01) & 0.54 (0.02) \\
Fashion-\textsc{mnist} & 0.43 (0.01) & 0.40 (0.01) & 0.40 (0.01) & 0.36 (0.03)  \\ \bottomrule
\end{tabular}
\caption{ARI means and standard deviations (within parentheses) associated with NNGP, random small submatrices, PAM and $k$-means. }
\label{realdatasets}
\end{table}

Both DPP approaches, based on NNGP and small submatrices, attain good results.
The NNGP approach outperformed all other three methods. Overall, the gains are
very important when compared with PAM and $k$-means.
The method based on random small submatrices also ourperformed the other methods,
except for the \textsc{mnist} dataset.

\section{Conclusions}\label{sec:conclusions}

Within the context of determinantal consensus clustering,
we proposed two different approaches to overcome the computational burden induced by the eigendecomposition of
the kernel matrix associated with large datasets.
The first one is based on NNGP, and the second one, is based on random sampling of small submatrices from the kernel matrix.

The NNGP approach finds a sparse approximation of the kernel matrix, based on nearest-neighbors.
The sparse matrix substitutes the original dense kernel matrix to extract the largest eigenvalues,
so as to be able to perform determinantal consensus clustering on the large dataset.
Simulations showed that using a high sparseness level and extracting a few largest eigenvalues
are enough to ensure good clustering results.
Extracting 1\% to 3\% of the eigenvalues
seems to be a reasonable strategy.
The amount of time needed to achieve the final clustering configuration is reduced considerably.

The approach based on random sampling of small submatrices from the kernel matrix is a good alternative to the NNGP approach.
Its performance is comparable to that of the NNGP approach, even though
the distribution of eigenvalues extracted with this method is not as close to the original eigenvalue distribution
of the kernel matrix.
Rather than keeping the original kernel matrix, this approach shows that considering the extraction of eigenvalues
from several rather small submatrices
of the kernel matrix is enough to obtain good results with consensus DPP.
Our simulations hint at sampling a number of small matrices in the range of $0.05n$ to $0.1n$.
If the submatrices are still too large, finding their sparse approximation with the $k$-nearest neighbors graph method is a good option, even with a small $k$.
In fact, for very large datasets, our simulations showed that it is strongly recommended to find a sparse approximation of the small submatrices. In order to speed up computational time, choosing a high sparseness level is the best option.

The two approaches are able to reach better quality results than consensus clustering applied to PAM.
Applications on large real datasets confirm the results found with the simulations.
Determinantal consesus clustering also proved to be superior to $k$-means for most of the datasets.
The presence of diversity in the sampled centroids helps to improve the quality of clustering and the stability of the results.

An issue we found with the NNGP approach, on some real datasets, is the possibility of
ill-conditioning of some of the intermediate calculation submatrices $W_{[1:i-1]}$ (see equation \eqref{eq:finley}).
This might occur because of strong similarity between some observations. As with the case of regression, the
original kernel matrix should be evaluated for numerical conditioning before attempting to use the NNGP approach.
A possible solution to avoid the problem is to eliminate observations that are too similar,
so as to only consider representative observations instead of all of them.
Another possibility is to regularize the submatrices by adding a small positive constant to the main diagonal,
as it is done in ridge regression.
Alternatively, the approach based on random small matrices might be used instead, if the problem arises.

\bibliographystyle{plain}
\bibliography{References2}  

\end{document}